\documentclass[twocolumn]{aastex63}

\usepackage{amsmath}
\usepackage{bbm,bm}
\usepackage{multirow}
\usepackage{colortbl}
\usepackage{hyperref}
\usepackage{outlines}
\usepackage{comment,color}

\graphicspath{{./}{figures/}}

\shorttitle{}
\shortauthors{}

\begin{document}

\title{Modeling Cosmic Ray Electron Spectra and Synchrotron Emission in the Multiphase ISM}

\author[0000-0001-8840-2538]{Nora B. Linzer}
\affiliation{Department of Astrophysical Sciences, Princeton University, 4 Ivy Lane, Princeton, NJ 08544, USA}
\email{nlinzer@princeton.edu}

\author[0000-0002-5708-1927]{Lucia Armillotta}
\affiliation{INAF Arcetri Astrophysical Observatory, Largo Enrico Fermi 5, Firenze, 50125, Italy}
\affiliation{Department of Astrophysical Sciences, Princeton University, 4 Ivy Lane, Princeton, NJ 08544, USA}

\author[0000-0002-0509-9113]{Eve C. Ostriker}
\affiliation{Department of Astrophysical Sciences, Princeton University, 4 Ivy Lane, Princeton, NJ 08544, USA}
\affiliation{Institute for Advanced Study, 1 Einstein Drive, Princeton, NJ 08540, USA}

\author[0000-0001-9185-5044]{Eliot Quataert}
\affiliation{Department of Astrophysical Sciences, Princeton University, 4 Ivy Lane, Princeton, NJ 08544, USA}

\begin{abstract}
We model the transport and spectral evolution of 1-100 GeV cosmic ray (CR) electrons (CREs) in TIGRESS MHD simulations of the magnetized, multiphase interstellar medium. We post-process a kpc-sized galactic disk patch representative of the solar neighborhood using a two-moment method for CR transport that includes advection, streaming, and diffusion. The diffusion coefficient is set by balancing wave growth via the CR streaming instability against wave damping (nonlinear Landau and ion-neutral collisions), depending on local gas and CR properties. Implemented energy loss mechanisms include synchrotron, inverse Compton, ionization, and bremsstrahlung. We evaluate CRE losses by different mechanisms as a function of energy and distance from the midplane, and compare loss timescales to transport and diffusion timescales. This comparison shows that CRE spectral steepening above $p=1$ GeV/c is due to a combination of energy-dependent transport and losses. Our evolved CRE spectra are consistent with direct observations in the solar neighborhood, with a spectral index that steepens from an injected value of -2.3 to an energy dependent value between -2.7 and -3.3. We also show that the steepening is independent of the injection spectrum. Finally, we present potential applications of our models, including to the production of synthetic synchrotron emission. Our simulations demonstrate that the CRE spectral slope can be accurately recovered from pairs of radio observations in the range 1.5-45 GHz.
\end{abstract}

\section{Introduction}\label{sec:intro}

Cosmic rays (CRs) are high energy, charged particles dominated by protons, with a smaller fraction of electrons, positrons, and heavier nuclei \citep[e.g.][]{blasi_origin_2013, zweibel_microphysics_2013, grenier_nine_2015}. Although there are a factor of $\sim10^9$ fewer CRs by number compared to thermal particles in the ambient interstellar medium (ISM), the CR pressure is similar to the thermal, turbulent, and magnetic components \citep[e.g.][]{2001RvMP...73.1031F,2004ARA&A..42..211E,2005ARA&A..43..337C}. This suggests that CRs can fundamentally impact the dynamics of gas in galaxies, perhaps most importantly by contributing to the driving of galactic winds \citep[e.g.][]{recchia_cosmic_2020, ruszkowski_cosmic_2023}. Moreover, CRs can also be an important source of heating and ionization of the dense ISM, which is shielded from far-ultraviolet (FUV) and photoionizing radiation \citep[e.g.][]{padovani_impact_2020, gabici_low_2022}. 

CRs are observed directly in the solar neighborhood using ground-based observatories or space-based detectors such as AMS-02 \citep{aguilar_first_2013} or Voyager \citep{stone_cosmic_2019}. These measurements show that above $\sim 1$ GeV, CRs follow a broken power law  distribution that extends up to PeV values. Lower energy (MeV-GeV) CRs are also present and are more important to ionization and heating of the ISM \citep[e.g.][]{draine_physics_2011}, but the spectrum is much more uncertain due to modulation by the Solar wind \citep[e.g.][]{padovani_impact_2020}. Beyond the local neighborhood, our understanding of CR properties and transport within the ISM depends on indirect probes of their spectral and spatial distribution. For example, hadronic interactions between CR protons and the ambient ISM produce neutral pions which decay and emit gamma radiation. Gamma-ray observations of high-density environments can therefore help to constrain the low energy CR spectrum \citep[e.g.][]{2017A&A...606A..22N}, and gamma-ray emission from galactic center regions can help to constrain CR transport under starburst conditions \citep[e.g.][]{2021MNRAS.502.1312C}. Gamma-ray radiation can also be produced through inverse Compton (IC) or bremsstrahlung interactions of CR leptons \citep[e.g.][]{zweibel_microphysics_2013}.

When gamma-ray observations are not possible, radio synchrotron emission can provide insight into the CR population. Synchrotron radiation is produced through the interaction of CR electrons (CREs) with the local magnetic field. Therefore, even though CREs make up a much smaller fraction of the total CR population, about a factor of 100 fewer in number compared to protons \citep[e.g.][]{zweibel_microphysics_2013}, they are important probes of the CR distribution in both the Milky Way and in extragalactic sources. Synchrotron observations are also commonly employed to obtain estimates of magnetic field strengths, under the assumption of equipartition between the CR and magnetic energy density \citep[e.g.][]{beck_revised_2005}. This assumption is not necessarily valid in all galaxies, but still provides one of the only ways to estimate the magnitude of the magnetic field in extragalactic sources. 

Inferring the distribution of CRs from synchrotron emission depends on accurately understanding the relative spectra of CR electrons and protons. This is not straightforward, since the CRE population has both a much lower amplitude compared to that of the protons, and also a different spectral shape. The distribution of CR protons above GeV energies is well approximated by a single power law \citep[e.g.][]{padovani_spectral_2021}. Electrons, however, are subject to significant energetic losses including ionization, bremsstrahlung, synchrotron and IC \citep[e.g.][]{schlickeiser_cosmic_2002}. These losses each have a different scaling with energy, thereby producing energy dependent steepening of the CRE spectrum \citep[e.g.][]{evoli_signature_2020}. 

With numerical models of the ISM that have realistic magnetic field structure (requiring both sheared rotation and turbulence as driven by correlated supernovae) and realistic structure of the multiphase gas, it is possible to directly simulate the transport of both CR protons and electrons. By modeling a range of CR energies, it is furthermore possible to investigate the energy dependence of CR transport including the effects of both energy-dependent scattering and losses. Simulations of this kind enable the creation of synthetic synchrotron emission, therefore providing a test of standard CR diagnostic techniques. In this work, we will focus on how multi-frequency synchrotron observations are employed to obtain the slope of the CRE spectrum as a function of energy.  

Previously, \cite{armillotta_cosmic-ray_2021, armillotta_cosmic-ray_2022,armillotta_cosmic-ray_2024} investigated the transport of GeV CR protons within the TIGRESS magnetohydrodynamic (MHD) simulations, which model local regions of  galactic disks \citep{kim_three-phase_2017, kim_first_2020}, allowing for a range of environments. Because the original TIGRESS simulations did not include CRs, \cite{armillotta_cosmic-ray_2021, armillotta_cosmic-ray_2022} calculated the transport of CRs in post-processing using the two-moment algorithm for CR transport \citep{jiang_new_2018} implemented in the \textit{Athena}++ MHD code \citep{stone_athena_2020}. In the original extension of the two-moment scheme described in \citet{armillotta_cosmic-ray_2021}, self-consistent scattering was implemented (see below) for a single relativistic fluid, considering either a CR kinetic energy of 1 GeV or 30 MeV. 

CR transport is controlled at large scales by the geometry of the magnetic field, which is advected by the thermal gas, and at small scales by energy-dependent scattering \citep[e.g.][]{ zweibel_basis_2017,blasi_spectral_2012, evoli_origin_2018}. Low and moderate-energy CRs ($\lesssim 100$~GeV), which resonate only with extremely small-scale perturbations in the magnetic field, are believed to scatter primarily off Alfvén waves excited by the CRs themselves via the resonant streaming instability \citep[the self-confinement scenario, e.g.][]{kulsrud_effect_1969, wentzel_cosmic-ray_1974}.  This is in contrast to higher-energy ($\gtrsim 100$~GeV) CRs, which are believed to be scattered by turbulent fluctuations that originate from other ISM dynamical processes \citep[the extrinsic turbulence scenario e.g.][]{chandran_confinement_2000,yan_scattering_2002}.

In line with the self-confinement scenario, \cite{armillotta_cosmic-ray_2021} treat the transport of the CR fluid in terms of advection along with the background thermal gas, streaming at the local Alfvén speed, and diffusion relative to the Alfvén wave frame. The rate of diffusion is determined by a scattering coefficient that depends on local gas properties, rather than using a constant value or a value that scales only with CR energy, as is commonly done in ISM simulations including CRs \citep[see reviews by][]{hanasz_simulations_2021, ruszkowski_cosmic_2023}. Specifically, in our approach the scattering coefficient is determined by the local balance between wave excitation and damping mediated by local gas properties (considering both ion-neutral damping and nonlinear Landau damping, e.g. \citealt{kulsrud_plasma_2005}). \cite{armillotta_cosmic-ray_2021} found the scattering coefficient to vary significantly between different ISM phases, spanning more than four orders of magnitude. To properly represent transport in a resolved, multiphase ISM, it is therefore critical to include this self-consistent computation of the diffusion rate, as CR transport differs greatly depending on local ISM properties including density, sound speed, and ionization fraction.  

For the present work, we extend the scheme of \cite{armillotta_cosmic-ray_2021} to track the propagation of spectrally resolved CR protons and electrons. The new scheme models the transport of multiple CR fluids, each representing CRs of a given species within a given range of momentum, and incorporates energy-dependent injection, scattering, and collisional losses. We use the updated scheme to calculate the transport of $1-100$~GeV CR protons and electrons by post-processing the TIGRESS simulation of the solar neighborhood environment. In this work, we make the simplifying approximation that CRs in different momentum bins  evolve independently. This method can be improved upon in the future to account for coupling between bins due to energy losses of individual particles, which can be implemented so as to be conservative  \citep[e.g.][]{girichidis2020spectrally}.  While the spectral decoupling of bins is a limitation, our model nevertheless enables a  study of energy-dependent transport of CREs in physically realistic models of the multiphase ISM at very high spatial resolution, in which the scattering rate is computed within the self-confinement paradigm and dynamical transport (both advection and streaming) are included.

Because the underlying TIGRESS ISM model is consistent with solar neighborhood conditions \citep[see][]{kim_three-phase_2017,Gong2018,Kado-Fong2020,kim_first_2020}, and we adopt a CR injection rate based on local constraints, our results can be compared to direct observations of CRs in the solar neighborhood. The synthetic synchrotron emission obtained from our simulations is at a spatial resolution of 8 pc, equivalent to an angular resolution of approximately 1.65'' for a galaxy at 1 Mpc. This resolution is achievable with observations from surveys such as the Local Group L-Band Survey \citep{lglbs} using the Karl G. Jansky Very Large Array (VLA) or future observations using the next-generation VLA \citep[e.g.][]{mckinnon_2019} or Square Kilometre Array \citep[e.g.][]{braun_2015}. In this paper, we focus on the analysis of CREs only, while the analysis of CR protons will be presented in a separate publication (Armillotta et al.\ 2025, accepted).

We note that other groups have performed similar studies of spectrally-resolved CRE transport in MHD simulations for comparison to observations, although these are generally based on larger scale simulations \citep[e.g.][]{werhahn_cosmic_2021a, hopkins_first_2022}. These recent works, as well as our approach, all build off of a large literature of CR transport studies. There are many code packages that have been developed  to solve for the propagation and spectral evolution of multiple CR species in static models of the Galaxy including GALPROP \citep{strong1998propagation,moskalenko1998production}, USINE \citep{maurin2001cosmic,putze2010markov}, DRAGON \citep{evoli2008cosmic, evoli2017cosmic, maccione2011dragon}, and PICARD \citep{kissmann2014picard}. The  methods adopted in these and similar traditional CR propagation packages are extremely effective in reproducing a wide range of CR observables, including spectra, abundances of different CR species, and non-thermal emission. They rely, however, on simplified prescriptions for the gas properties and CR propagation. In particular, since the underlying gas models are not based on direct MHD simulations, they do not include dynamical transport of CRs  based on a realistic turbulent magnetic field, and the scattering coefficients are set as model parameters rather than being computed locally from spatially-varying ISM properties. Codes such as CREST \citep{winner2019evolution, winner2020evolution} and CRIPTIC \citep{krumholz2022cosmic,Sampson2023} model CR transport by solving the Fokker-Planck equation (with some approximations), but can be applied to outputs from MHD simulations, allowing for a more realistic magnetic field structure. Additionally, models which self-consistently evolve spectrally-resolved CREs and MHD variables in a time-dependent fashion have been used to study many different astrophysical environments across a wide range of scales including galaxy clusters \citep[e.g.][]{miniati2001cosmic}, the CGM and ISM \citep[e.g.][]{girichidis2020spectrally, girichidis_spectrally_2022, girichidis2024spectrally, ogrodnik_implementation_2021}, Fermi bubbles \citep[e.g.][]{yang2017spatially, boss2023crescendo}, and MHD flows and shocks \citep[e.g.][]{vaidya2018particle}. A wide range of numerical approaches based on different approximations have been adopted for these studies; we refer the reader to the original publications for details.

The paper is organized as follows. In \autoref{sec:methods}, we describe the TIGRESS simulation of the solar neighborhood, summarize the CR transport method used in \cite{armillotta_cosmic-ray_2021}, and describe the updates made to model multi-group CR spectra. In \autoref{sec:results}, we present the first results of our multi-group simulations, including the spatial distribution (\autoref{sec:results_dist}) and energy spectrum (\autoref{sec:spectrum_result}) of CREs. \autoref{sec:synch_results} presents an application of our model to testing diagnostic methods, in which we compare the CRE spectrum that would be inferred from synchrotron observations to the underlying CRE spectrum, as would be measured with direct detection. We compare our results to other recent models and discuss further connections to observations in \autoref{sec:disc}. A summary and prospectus of future applications is provided in \autoref{sec:conc}.

\section{Methods}\label{sec:methods}

We model the propagation of CRs within the TIGRESS simulation of the solar neighborhood (see \autoref{sec:tigress}) using a two-moment scheme for CR transport, with detailed methods developed in \citet{armillotta_cosmic-ray_2021,armillotta_cosmic-ray_2024} described in \autoref{sec:transport}. Extension of the CR method to model separate groups of CR protons and electrons from 1-100 GeV, with a defined injection spectrum and energy-dependent scattering driven by the streaming instability, is described in \autoref{sec:spectrum} - \autoref{sec:scattering}. \autoref{sec:ion_frac} - \autoref{sec:synch} describe the methods implemented to model losses and to compute synchrotron emission.  

\subsection{TIGRESS Simulation}\label{sec:tigress}

The TIGRESS framework \citep{kim_three-phase_2017} is built with the code \textit{Athena} \citep{stone_athena_2008} in which the ideal MHD equations are solved in a shearing periodic box \citep{stone_implementation_2010}. TIGRESS is designed to model the multiphase ISM by including star formation feedback via time-dependent FUV radiation associated with recent star formation, which heats warm and cold gas via the photoelectric effect on small grains, as well as Type II supernova (SN) energy inputs, which create the hot medium via strong shocks. TIGRESS includes optically thin cooling, taking a simplified approach with a fitting function for warm/cold gas and tabulated values for hot gas in the original implementation of \citet{kim_three-phase_2017} for the models employed here (a more sophisticated non-equilibrium formulation was introduced by \citealt{kim_photochemistry_2023}). Gravity is included both as gas self-gravity, which leads to localized collapse and creation of star cluster particles, and an external potential from both a stellar disk and dark matter halo. The simulation accurately evolves magnetic fields, which is necessary for both CR transport and synchrotron emission.

Star formation is modeled through the use of sink particles, which represent unresolved stellar clusters. Based on the age of each particle, and assuming a stellar population with a Kroupa IMF, the rate of SNe and FUV luminosity is taken from STARBURST99 \citep{1999ApJS..123....3L}. In the CR post-processing, star cluster particles act as sources of CR injection (see \autoref{sec:injection}). 

\cite{kim_first_2020} and \citet{2022ApJ...936..137O} present results from a range of TIGRESS models covering different galactic conditions, and \cite{armillotta_cosmic-ray_2022} studied the transport of 1 GeV CR protons in three different environments via post-processing. Here, we apply CR post-processing to the {\tt R8} simulation, which is designed to represent conditions similar to the solar neighborhood. The same MHD model was previously used for the analysis of GeV CRs in \cite{armillotta_cosmic-ray_2021} and \citet{armillotta_cosmic-ray_2024}. The {\tt R8} model represents a patch of the galactic disk with $L_x = L_y = 1024$ pc and height $L_z = 7168$ pc. The simulation we use has a uniform grid with $\Delta x = 8$ pc. Although there are TIGRESS simulations with higher resolution, the models at a resolution of 8 pc show convergence of both gas and CR properties \citep{kim_three-phase_2017, armillotta_cosmic-ray_2021}.

For this work, we select eight snapshots from the {\tt R8} simulation between times of $200-550$ Myr. The snapshots were chosen to exclude the initial transient behavior of the TIGRESS simulations and to cover a range of star formation and feedback cycles. 

\subsection{Cosmic Ray Transport Equations}
\label{sec:transport}

We model CR transport with the two-moment scheme originally developed by \cite{jiang_new_2018} within the code \textit{Athena}++ \citep{stone_athena_2020}. This method was later adapted for use with the TIGRESS framework by \cite{armillotta_cosmic-ray_2021}. Extending the previous work, which considered only one CR fluid representing CR protons with a single energy, here we evolve multiple non-interacting CR fluids, each representing either CR protons or electrons within a given range of energies. Each individual CR component, denoted by the subscript $j$, is evolved independently, with transport following the same equations as in \cite{armillotta_cosmic-ray_2021}:

\begin{multline}
    \frac{\partial e_{\textrm{c},j}}{\partial t} + \nabla \cdot \mathbf{F}_{\textrm{c},j} = \\ - (\mathbf{v} + \mathbf{v}_{\textrm{s},j}) \;\cdot \stackrel{\leftrightarrow}{\sigma}_{\textrm{tot,}j} \cdot \; [\mathbf{F}_{\textrm{c},j} - \mathbf{v} \cdot (\stackrel{\leftrightarrow}{\mathbf{P}}_{\textrm{c},j} + \textrm{$\textit{e}_{\textrm{c},j}$}\stackrel{\leftrightarrow}{\mathbf{I}})]
    \label{eq:energy}
\end{multline}
\begin{multline}
    \frac{1}{v_\textrm{m}^2}\frac{\partial \mathbf{F}_{\textrm{c},j}}{\partial t} + \nabla \cdot \stackrel{\leftrightarrow}{\textbf{P}}_{\textrm{c},j} = \\ -\stackrel{\leftrightarrow}{\sigma}_{\textrm{tot},j} \cdot \; [\mathbf{F}_{\textrm{c},j} - \mathbf{v} \cdot (\stackrel{\leftrightarrow}{\textbf{P}}_{\textrm{c},j} + \textrm{$\textit{e}_{\textrm{c},j}$}\stackrel{\leftrightarrow}{\mathbf{I}})]\,,
    \label{eq:flux}
\end{multline}
where $e_{\textrm{c},j}$ is the CR energy density, $\textbf{F}_{\textrm{c},j}$ is the CR flux vector, and $\stackrel{\leftrightarrow}{\textbf{P}}_{\textrm{c},j}$ is the CR pressure tensor. We adopt an adiabatic equation of state and assume the CR pressure is isotropic, such that $\stackrel{\leftrightarrow}{\textbf{P}}_{\textrm{c},j} = P_{\textrm{c},j}\stackrel{\leftrightarrow}{\bf{I}}$, with $P_{\textrm{c},j} = (\gamma - 1)e_{\textrm{c},j}$. Here, $\stackrel{\leftrightarrow}{\bf{I}}$ is the identity tensor. 

In line with most MHD simulations that include a CR fluid, we make the simplifying assumption that CRs are relativistic, and therefore, the adiabatic index of the CR fluid is $\gamma_{ad} = 4/3$ \citep[e.g.][]{pfrommer_simulating_2017, butsky_impact_2020, bustard_cosmicray_2021, werhahn_cosmic_2021b}. We note that while this assumption is always valid for CR electrons, the CR protons modeled in this work are not all strictly relativistic, so the actual adiabatic index of the CR proton fluids can differ slightly from $4/3$. Specifically, the true value of $\gamma_{ad}$ is $\lesssim 1.4$ for $p=2$~GeV/c (corresponding to our lowest momentum bin), and it decreases with increasing $p$, reaching $\gamma_{ad} = 4/3$ for $p \geq 10$~GeV/c \citep[see][]{girichidis_spectrally_2022}.

Theoretically, the maximum velocity at which relativistic CRs can propagate is the speed of light, $c$. However, in \autoref{eq:flux}, we limit the maximum CR velocity to $v_\mathrm{m} \approx 10^4 \textrm{ km/s} \ll c$. \cite{jiang_new_2018} show that the simulation outcomes are not sensitive to the value of $v_\mathrm{m}$ as long as $v_\mathrm{m}$ remains larger than any other speed in the simulation. The reduced speed of light approximation allows us to reduce the computational time of the simulation by increasing the size of the time step allowed by the CFL condition. 

In \autoref{eq:energy} and \autoref{eq:flux}, $\textbf{v}$ is the gas velocity, while $\mathbf{v}_{\textrm{s},j}$ is the streaming velocity in the direction determined by the CR pressure gradient:
\begin{equation}
    \textbf{v}_{\textrm{s},j} = - v_{\textrm{A},i} \hat{B}\frac{\hat{B} \cdot \nabla P_{\textrm{c},j}}{|\hat{B} \cdot \nabla P_{\textrm{c},j}|}\,.
\end{equation}
Here, $v_{\textrm{A},i} = B/\sqrt{4 \pi \rho_{\mathrm{i}}}$ is the ion Alfvén speed, with $B$ the magnetic field strength and $\rho_{\mathrm{i}}$ the ion density, allowing for partial ionization in gas at $T<5\times10^4$ K (see \autoref{sec:ion_frac}). The magnitude of the CR streaming velocity is the same for all CR components, but the direction differs depending on the value of $\nabla P_{\textrm{c},j}$. 

The diagonal tensor $\stackrel{\leftrightarrow}{\sigma}_{\mathrm{tot},j}$ is the wave-particle interaction coefficient. In the \citet{jiang_new_2018} implementation, its component along the magnetic field direction accounts for both scattering and streaming, and is defined as: 
\begin{equation}
    \sigma_{\textrm{tot}, \parallel, j}^{-1} = \sigma_{\parallel, j}^{-1} + \frac{v_{\textrm{A},i}}{|\hat{B} \cdot \nabla P_{\textrm{c},j}|}(P_{\textrm{c},j} + e_{\textrm{c},j})\,,
\end{equation}
where $\sigma_{\parallel,j}$ is the scattering coefficient (see \autoref{sec:scattering}). Perpendicular to the magnetic field direction, there is only scattering, such that  $\sigma_{\textrm{tot},\perp,j} = \sigma_{\perp,j}$. As noted by \cite{armillotta_cosmic-ray_2021}, in the time-independent limit of \autoref{eq:flux} ($\partial \mathbf{F}_{\textrm{c},j}/\partial t \approx 0$ or large $v_\mathrm{m}$), the flux becomes:
\begin{equation}
    {\bf{F}}_{\textrm{c},j} = \dfrac{4}{3}  e_{\textrm{c},j} ({\bf{v}} + {\bf{v}}_{\textrm{s},j} ) - {\stackrel{\leftrightarrow}{\sigma_{j}}}^{-1} \cdot \nabla P_{\textrm{c},j}
    \label{eq:steadyflux}
\end{equation}
and the work on the right-hand side of the CR energy equation reduces to 
\begin{equation}
(\mathbf{v} + \mathbf{v}_{\mathrm{s},j})\cdot \nabla P_{\mathrm{c},j}
\end{equation}
This clearly shows that the transport of CRs is given as a sum of advection (${4}/{3}  e_{\textrm{c},j} {\bf{v}}$), streaming (${4}/{3}  e_{\textrm{c},j} {\bf{v}}_{\textrm{s},j}$), and diffusion ($- {\stackrel{\leftrightarrow}{\sigma_{j}}}^{-1} \cdot \nabla P_{\textrm{c},j}$).

We note that, by evolving each CR momentum (or energy) component 
independently, our model neglects transfer across momentum bins. In Armillotta et al.\ (2025, accepted), we discuss how \autoref{eq:energy} and \autoref{eq:flux} would be modified when adiabatic energy transfer is taken into account, showing that the coefficients of the source terms change. Specifically, they become $-(\gamma_\mathrm{j} - 3) \stackrel{\leftrightarrow}{\sigma}_{\mathrm{tot},j} \cdot (\mathbf{v} + \mathbf{v}_\mathrm{s, \mathit{j}}) \cdot (\mathbf{F}_{\mathrm{c},j} - (\gamma_j / 3) \mathbf{v} {e}_{c,\mathit{j}})$ and 
$-\stackrel{\leftrightarrow}{\sigma}_{\mathrm{tot},j} \cdot (\mathbf{F}_{\mathrm{c},j} - (\gamma_j / 3) \mathbf{v} {e}_{c,\mathit{j}})$ for \autoref{eq:energy} and \autoref{eq:flux}, respectively, assuming that the CR distribution function in each bin can be approximated by a power law $f_j \propto p^{-\gamma_j}$. For $\gamma_j = 4$, we recover the exact form of the source terms used in \autoref{eq:energy} and \autoref{eq:flux}. In our simulations, $\gamma_j$ varies from $-4.3$ (the injection slope) to $-5.3$ (see \autoref{sec:results}), implying that the resulting difference in the individual source terms is small. As demonstrated in Armillotta et al.\ (2025, accepted), while accounting for energy transfer may lead to a slight shift of the final CR spectrum in momentum space, the overall  behavior and key conclusions of the present study remain unchanged.

If we included energy transfer between momentum bins, there would be an additional source term associated with CR energetic losses (described in \autoref{sec:losses}). When any CR interacts with the background gas, radiation field, or magnetic field, it loses energy and would therefore move into a lower-momentum bin, but we do not add energy into the lower-momentum bin representing this shift. We find, however, that the injected CR spectrum is sufficiently steep that energy shifted from a higher to lower momentum bin is not significant.

Energy and momentum are transferred between the CRs and the surrounding gas at a rate determined by the right-hand side of  \autoref{eq:energy} and \autoref{eq:flux} respectively. Any energy and momentum lost (or gained) by the CRs is correspondingly applied as a source (or sink) term in the equations for gas energy and momentum. For the majority of the CR post-processing, however, we freeze the background gas and magnetic field distribution, and therefore do not include these additional terms in the MHD equations. We also omit the adiabatic work term from the CR energy equation, because it would otherwise introduce spurious sources of CR energy at interfaces between cold/warm and hot gas when there is no CR backreaction on the gas (see below).

After the total CR energy density has reached a steady state (e.g.\ $\int_\mathrm{Vol} \dot{e}_\mathrm{c} dx^3 \approx 0$, with \textit{Vol} the simulation volume), however, we let the gas and CRs evolve together for a short period (2 Myr) of MHD relaxation. \cite{armillotta_cosmic-ray_2024} found that when the gas is frozen, there may be locations (especially at shock interfaces between diffuse hot gas and other phases) where the magnetic fields remain aligned perpendicular to the density gradient.  In this situation, CRs may be trapped in dense regions, which produces large CR pressure gradients at the boundaries between the warm, dense gas and the hot, diffuse regions. If MHD back-reaction is not permitted, the large CR gradient at the interface is preserved.  As discussed in \citet{armillotta_cosmic-ray_2024}, it is therefore necessary during the stage of the solution when MHD is frozen to omit the work term in the CR energy equation to avoid an unphysical enhancement in CR energy. Conversely, we include the adiabatic work term during the MHD relaxation step. We note, however, that in the midplane region the work term is generally small compared to loss terms because the CR pressure is quite uniform, due to strong wave damping in neutral gas.

When the background gas is briefly allowed to evolve, the CR pressure gradient at interfaces causes expansion at the surface of dense regions. The magnetic and velocity fields reorient, becoming more in line with the CR pressure gradients. This allows the CRs to escape from the dense gas, such that the overall CR distribution becomes smoother. The orientations of the magnetic and velocity fields converge for evolution times $\gtrsim 1$ Myr, as was previously demonstrated in \citet{armillotta_cosmic-ray_2024}. Because star formation and feedback are not included during the MHD relaxation step, we cannot allow the gas to evolve for more than 2 Myr without losing a significant fraction of hot gas. Additionally, to be consistent with the lack of thermal energy and momentum injection, we do not inject CRs during this stage. 

Due to the relatively brief interval of MHD relaxation, the vertical CR energy density profiles do not evolve to an entirely new steady state. We find, however, that neither the vertical CR energy profile nor the CR energy spectrum change significantly before and after the MHD relaxation step. This is shown in \autoref{sec:app_mhd}.\footnote{A potential concern is that in a fully self-consistent CR-MHD simulation evolved over an extended period, both the vertical MHD and CR profiles could differ from those under our present approach due to CR pressure forces on the gas. However, preliminary analysis of a new fully self-consistent TIGRESS CR-MHD simulation with a single GeV CR component (C.-G. Kim et al, in prep.) shows that the CR pressure throughout the midplane region (at $|z|\lesssim 300$pc) is nearly the same as in the present simulations.} All results presented in this paper are taken at the end of the MHD relaxation step.

Other works including a variable scattering model based on self-confinement have found that while the midplane ISM structure does not change significantly with the addition of CR feedback -- due to efficient diffusion and thus negligible CR pressure gradients -- the extraplanar region is more strongly affected \citep[e.g.][]{sike2024cosmic, thomas2024}. Therefore, while we discuss the CR distribution in the extraplanar region qualitatively throughout \autoref{sec:results}, we note that these results may differ in fully self-consistent CR-MHD simulations. When comparing to to direct observations, we consider only the midplane ISM.

In \autoref{sec:injection} and \autoref{sec:losses}, we describe additional source and sink terms that enter the right-hand sides of \autoref{eq:energy} and \autoref{eq:flux}. These include the injection of energy into CRs by SNe, as well as various collisional processes through which CRs lose energy and momentum.

\subsection{CR Proton and Electron Momentum Spectra}
\label{sec:spectrum}

We model the CR spectrum by evolving 10 CR components -- 5 representing protons and 5 representing electrons -- each corresponding to a specific bin of momentum values. For the protons, the bins are equally divided in log space between momenta of $2-101$~GeV/c, corresponding to kinetic energies between $1-100$~GeV. The 5 bins are centered at $p_{j} = 2$, 5, 13, 36, and 101 GeV/c. Each bin has a width of $d\textrm{ln}p$ = 0.1. 

The momenta of the electron bins are based on the predictions of the self-confinement scenario, which we assume in this work. According to detailed modeling of the local CR spectra, CRs in the energy range investigated here ($1-100$~GeV) are mostly scattered by self-excited resonant Alfvén waves \citep[e.g.][]{zweibel_microphysics_2013}. Since the total CR energy density is dominated by protons, the scattering of the electrons is determined primarily by interactions with Alfvén waves generated by these protons. Therefore, we need to determine which electrons interact with the waves excited by the protons in any given momentum bin. The resonance condition of a CR with an Alfvén wave is given by
\begin{equation}
    \pm \Omega \approx -k v_{c,\parallel}\,,
\end{equation}
\citep[e.g.][]{bai_magnetohydrodynamic_2019}, where $k$ is the wavevector of the Alfvén wave and $\Omega = {q B}/({\gamma m c})$ is the gyrofrequency for a particle with charge $q$ (equal to $\pm e$  for protons or electrons), Lorentz factor $\gamma$, and particle mass $m$ in a magnetic field with magnitude $B$. The CR velocity $v_c$ is related to CR energy by  
\begin{equation}
    v_c = c\sqrt{1-\frac{1}{\gamma^2}}= \sqrt{1 - \bigg(\frac{m c^2}{E}\bigg)^2}
    \label{eq:v_c}
\end{equation}
for a CR with mass $m$ and relativistic energy $E$, and $v_{c,\parallel}$ represents the component of this velocity parallel to the background magnetic field.

For a CR proton and electron to interact with the same Alfvén wave of a given wavevector $k$, it must be true that $\Omega_{\rm p}/v_{\rm p} = \Omega_{\rm e}/v_{\rm e}$, where the subscripts $p$ and $e$ indicate protons and electrons, respectively. Substituting for the gyrofrequency, we find that $\gamma_p m_p v_p = \gamma_e m_e v_e$. CR protons and electrons that are resonant with the same Alfvén waves must have the same relativistic momentum. Therefore, to model both species, we use the same bins in momentum space for both protons and electrons and will consider protons and electrons with the same momentum to have the same scattering coefficient (see \autoref{sec:scattering}). 

As mentioned in \autoref{sec:transport}, each CR fluid component has an associated energy density and flux which are evolved according to \autoref{eq:energy} and \autoref{eq:flux} respectively. The fluids do not interact with each other except for the determination of the scattering coefficient. We note that the energy density of CRs within a given momentum bin is related to the CR distribution function $f(p)$ as follows:
\begin{equation}
\begin{split}
    e_{\textrm{c},j} &  = 4 \pi \int_\mathrm{p_{-}}^\mathrm{p_{+}} f(p) E(p) p^2 dp\\
    & \approx 4 \pi f(p_j) E(p_j) p_j^3 d\mathrm{ln}{p} \,,
    \label{eq:endens}
\end{split}    
\end{equation}
where $p_{\pm}$ are the extrema of the bin, and $E(p)$ is the total relativistic energy for a particle of relativistic momentum $p_j$. Taking the  approximation in the second line of \autoref{eq:endens} makes a negligible difference, changing $e_{\textrm{c},j}$ by less than 0.05\%, because our spectral bins are sufficiently small.

The initial distribution function is defined as $f(p) = C p^{-\gamma_{\rm input}}$ where $\gamma_{\rm input}$ is the initial spectral power law slope and $C$ is a normalization constant, and this power law is assumed to hold for $p \ge p_\mathrm{min}$. We choose $\gamma_{\rm input}$ to be consistent with the initial CR distribution produced in SNR, estimated by direct observations to be approximately $4.3$ \citep[see e.g.][]{caprioli_particle_2023}, although we will consider variations in this value.

We set the normalization of the CR proton distribution as follows. First, we simulate CR transport using the one-bin model as in \cite{armillotta_cosmic-ray_2021}. In this case, the full CR population is modeled as if all CRs were protons with kinetic energy of 1 GeV. When this model approaches steady state, we use the resulting proton energy density ($e_{\textrm{c,p,tot}}$) to solve for $C$ by setting
\begin{equation}
  e_{\textrm{c,p,tot}} = 4 \pi \int_{\textrm{p$_\textrm{min}$}}^\infty f(p) E(p) p^2 dp  
  \label{eq:tot_ec}
\end{equation}
where $p_\textrm{min}$ = 1 GeV/c. Below this value the injection slope is uncertain, and the relativistic assumption is not valid. By changing the value of $p_\textrm{min}$, we only change the overall normalization of the initial CR distribution. The CRE spectrum is initialized such that it has the same spectral shape as the CR protons, but $e_{\textrm{c,e,tot}}$ is reduced by a factor of 50 compared to $e_{\textrm{c,p,tot}}$. This is consistent with direct observations of the CR spectra \citep[e.g.][]{zweibel_microphysics_2013}. 

This process for setting the normalization of the energy spectrum of the CR protons and electrons in the initial conditions is done only to reduce the time it takes for the multi-fluid simulations to reach a steady state. The final CR distributions are not dependent on the initialization.

\subsection{CR Energy Injection}
\label{sec:injection}

The method for injecting energy of the CR fluid onto the grid based on SN feedback from star cluster particles is described in detail by \cite{armillotta_cosmic-ray_2021} when considering CRs with a single energy. We extend this method to allow for injection of CRs with different momenta. The total rate of CR energy injected from SNe is 
\begin{equation}
\dot{E_c} = \epsilon_{\rm inj, tot}E_{\rm SN}\dot{N}_{\rm SN}
\label{eq:energy_inj}
\end{equation}
where $E_{\rm SN} = 10^{51}$ erg is the energy released by one SN event, and $\dot{N}_{\rm SN}$ is the rate at which SNe occur. The SN rate from a given star cluster particle is defined as $\dot{N}_{\rm SN} = m_{\textrm{sp}}\xi_{\textrm{SN}}(t_{\textrm{sp}})$ where $m_{\textrm{sp}}$ is the star cluster particle mass, and $\xi_{\textrm{SN}}(t_{\textrm{sp}})$ is the rate of SNe per unit mass for a star cluster particle of age $t_{\textrm{sp}}$ as computed using STARBURST99 \citep{kim_three-phase_2017}. 

In the energy injection expression, $\epsilon_{\rm inj, tot}$ represents the fraction of SN energy that goes into production of CRs, assumed to be equal to 0.1 \citep[see e.g.][]{caprioli_particle_2023}. When considering only one CR fluid, \cite{armillotta_cosmic-ray_2021} assume the total injected energy goes into acceleration of 1 GeV CRs. To instead model a CR spectrum, we must divide the total injected energy between CRs of different momenta. We do so as follows.

The total CR energy density produced by one SN is
\begin{equation}
    e_{\textrm{inj,tot}} = \epsilon_{\textrm{inj,tot}} \frac{E_{\rm{SN}}}{V}\,,
    \label{eq:einjtot}
\end{equation}
with $V$ the volume in which the energy is injected. The injected CR energy density, $e_{\textrm{inj,tot}}$, is determined by \autoref{eq:tot_ec} where the distribution function $f(p)$ is replaced by the distribution of injected particles with the same power-law index ($\gamma_{\rm inj} = -4.3$) but a different normalization.\footnote{More generally, $f(p)$ is often treated as a piecewise power law.  Our implicit assumption in adopting a single power law with $p_\mathrm{min}=1\;\mathrm{GeV}/c$ is that $f(p)$ is flat enough below 1 GeV that the total CR energy at $p<$1 GeV is negligible.} 

The energy density injected into a single bin is
\begin{equation}
    e_{\textrm{inj},j} = \epsilon_{\textrm{inj},j} \frac{E_{\rm{SN}}}{V} \,
    \label{eq:einjbin} 
\end{equation}
where $\epsilon_{\textrm{inj,j}}$ is the fraction of the SN energy injected into the $j$th bin. We substitute $E_\mathrm{SN}/V$ from \autoref{eq:einjtot} into \autoref{eq:einjbin}. Then, using \autoref{eq:tot_ec} and the approximation in \autoref{eq:endens}, we obtain
\begin{equation}
\epsilon_{\textrm{inj},j} = \epsilon_{\textrm{inj,tot}} \frac{p_j^{-\gamma_{\rm inj}} E(p_j) p_j^3 d\textrm{ln}p}{\int_{\textrm{p$_\textrm{min}$}}^\infty p^{-\gamma_{\rm inj}} E(p) p^2 dp}\,.
    \label{eq:eps_inj}
  \end{equation}

Taking a relativistic limit $E(p)\approx p c$ for \autoref{eq:eps_inj}, this simplifies to
\begin{equation}
\epsilon_{\textrm{inj},j} \approx \epsilon_{\textrm{inj,tot}} (\gamma_{\rm inj} -4)\left(\frac{p_\textrm{min}}{p_j}\right)^{\gamma_{\rm inj} -4}  d\textrm{ln}p \,,
    \label{eq:bin_inject}
\end{equation}
(assuming $\gamma_{\rm inj}>$ 4).

As in \autoref{sec:spectrum}, we inject electrons with the same spectral slope as the protons, but change the normalization such that the magnitude of the electron distribution is reduced by a factor of 50 compared to the protons ($\epsilon_{\rm inj, tot} = 0.002$). Other simulations of multiple CR species have also adopted an injected ratio of 0.02 between electrons and protons \citep[e.g.][]{werhahn_cosmic_2021b}.

Like in \cite{armillotta_cosmic-ray_2021}, the injected energy appears as a source term in \autoref{eq:energy}. The injection follows a Gaussian distribution around each star cluster particle, such that the total source term in a cell at location $\mathbf{x}$ is
\begin{equation}
    Q(\mathbf{x}) = 
    \frac{1}{(2 \pi \sigma_\mathrm{inj}^2)^{3/2}}
    \sum_{\rm{sp}=1}^{N_{\rm{sp}}} \dot{E}_{\textrm{c,sp}}\cdot \rm{e}^{-r_{\rm{sp}}^2/2\sigma_{\rm{inj}}^2}
\end{equation}
where $Q$ represents the injected CR energy density per unit time and $\dot{E}_{\textrm{c,sp}}$ is the energy injection rate (\autoref{eq:energy_inj}) from one star cluster particle, with $r_\mathrm{sp}=|\mathbf{x} - \mathbf{x}_\mathrm{sp}|$ the distance from the cell to each contributing star particle. The sum is taken over all star particles in the simulation with age less than 40 Myr. We adopt $\sigma_{\rm inj}=32$ pc; \cite{armillotta_cosmic-ray_2021} found that the choice of $\sigma_{\rm inj}$ did not have a significant impact on the final CR distribution.

We note that in \autoref{eq:energy}, the CR energy density is approximately linearly proportional to Q with the exception of $\sigma_{\rm tot}$ which has a weak dependence on $e_c$ (see \autoref{sec:scattering}). Therefore, the overall normalization of the injected CR spectrum (see \autoref{eq:endens} and \autoref{eq:tot_ec} in \autoref{sec:spectrum}) does not have a significant effect on the rate of CR transport. The final CR distribution (and resulting synchrotron emissivity) can be renormalized after it has evolved to a steady state. This would represent, for example, a different choice of $p_\textrm{min}$ or $\epsilon_{\textrm{inj,tot}}$.

\subsection{Scattering Coefficient}
\label{sec:scattering}

We compute the scattering coefficient in line with the predictions of the self-confinement picture. Similar to \cite{armillotta_cosmic-ray_2021}, the scattering coefficient, $\sigma_\parallel$, is determined by balancing the growth of Alfvén waves with different damping mechanisms. We will provide a brief summary of the derivation here as well as a description of how we include the contribution of CREs.

\cite{armillotta_cosmic-ray_2021} demonstrated that, for quasi-steady CR flux (see \autoref{eq:steadyflux}), the growth of Alfvén waves due to the streaming instability \citep[e.g.][]{kulsrud_plasma_2005} can be written as:
\begin{equation}
    \Gamma_{\textrm{stream}}(p_{1,j}) = \frac{\pi^2}{4}\frac{\Omega_0 m v_{\textrm{A,i}}}{B^2}\frac{|\hat{B} \cdot \nabla P_{\textrm{c},j}|}{\sigma_{\parallel,j}P_{\textrm{c},j}}n_{1,j}\,,
    \label{eq:growth}
\end{equation}
where $\Omega_0 m = e B /  c$, with $\Omega_0$ the cyclotron frequency, and $p_{1,j} =  \Omega_0 m/ k_{\rm j}=e B/(k_j c)$ the resonant momentum for wave number $k_{ j}$, which we set to the momentum of the CR bin (see \autoref{sec:spectrum}). The waves propagate at the ion Alfvén speed, $v_{\textrm{A,i}} = B / \sqrt{4 \pi \rho_{\mathrm{i}}}$, where $\rho_{\mathrm{i}}$ is the ion density as defined in \autoref{sec:ion_frac}. The value of $n_{1,j}$ is defined as 
\begin{equation}
    n_{1,j} \equiv 4 \pi p_{1,j} \int_{p_{1,j}}^{\infty} p f(p) dp\,.
    \label{eq:n1_0}
\end{equation}
Here, we assume $f(p) = C (p/p_{1,j})^{-\gamma_{\rm obs}}$, where C is a normalization constant and $\gamma_{\rm obs} = 4.7$ represents a fixed power law slope consistent with observations of CR protons with kinetic energy $\gtrsim 1$~GeV \citep[e.g.][]{grenier_nine_2015}. The normalization constant is computed from the approximation in \autoref{eq:endens}, using the energy density of CRs within the bin centered at $p_{1,j}$, i.e.\ $C = e_{c}(p_{1,j})/[4 \pi E(p_{1,j})p_{1,j}^3 d\textrm{ln}p]$. Therefore, $n_{1,j}$ simplifies to
\begin{equation}
    n_{1,j} =  \frac{e_{c}(p_{1,j})}{2.7 E(p_{1,j}) d\textrm{ln}p}\,
    \label{eq:n1}\,.
\end{equation}
We note that properly speaking, the factor 2.7 in the denominator would be replaced by a numerical factor $\gamma_j -2$ that depends on the local slope of the CR spectrum.  Since the final, evolved slopes we obtain are very close to observed values, however, we have adopted the above approach for simplicity.  

The value of $n_{1,j}$ is only dependent on the CR energy density and relativistic energy of bin $j$. We note that because $e_\mathrm{c}/E(p) \propto p^3 f(p)$, $n_{1,j}$ decreases with increasing momentum $p$ for distribution functions with slope less than $-3$, which is always true in our simulations.

The growth of Alfvén waves is balanced by damping mediated by local gas properties. We consider both ion-neutral (IN) and nonlinear Landau (NLL) damping. The IN damping is due to collisions between ions and neutrals, and is the dominant damping mechanism in denser, poorly-ionized gas, where neutrals are not tied to magnetic fields. The IN damping rate is \citep{kulsrud_effect_1969}
\begin{equation}
\Gamma_{\textrm{damp,in}} = \frac{1}{2}\frac{n_{\mathrm{n}} m_{\mathrm{n}}}{m_{\mathrm{n}} + m_{\mathrm{i}}}\langle \sigma v \rangle_\textrm{in}\,,
\label{eq:damp_in}
\end{equation}
where $n_{\mathrm{n}}$ and $m_{\mathrm{n}}$ are the number density and mean mass of neutral gas particles, and $m_{\mathrm{i}}$ is the mean mass of ions (see \autoref{sec:ion_frac} for the definitions of these values). $\langle \sigma v \rangle_\textrm{in} \sim 3 \times 10^{-9}\textrm{ cm}^3\textrm{ s}^{-1}$ is the rate coefficient for IN collisions \citep{draine_physics_2011}. If we assume that only protons contribute to the Alfvén wave growth, and balance the wave growth rate and the IN damping rate, $\Gamma_{\textrm{stream}} = \Gamma_{\textrm{damp,in}}$, solving for $\sigma_\parallel$ we find 
\begin{multline}
\sigma_{\parallel\textrm{,in}}(p_{1,j}) = \\ \frac{\pi}{8}\frac{|\hat{B} \cdot \nabla P_{\textrm{c},j}|}{v_{\textrm{A}, i}P_{\textrm{c},j}}\frac{\Omega_0 m}{n_\textrm{n}\langle\sigma v \rangle_\textrm{in}}\frac{(m_\textrm{n} + m_{\mathrm{i}})}{m_{\mathrm{i}} m_\textrm{n}}\frac{n_{1,j}}{n_{\mathrm{i}}}\,.
\end{multline}

This is the form of the scattering coefficient used in \cite{armillotta_cosmic-ray_2021}. We now consider the contribution of both protons and electrons, so we instead solve $\Gamma_{{\rm stream, p}} + \Gamma_{{\rm stream, e}} = \Gamma_{{\rm damp}}$. Because this is a linear equation in $\sigma_\parallel$, the total value of $\sigma_\parallel$, including contributions from both protons and electrons, is simply
\begin{multline} \sigma_{\parallel\textrm{,in,tot},j} = \frac{\pi}{8}\frac{\Omega_0 m}{v_{\textrm{A}, i} n_\textrm{n}\langle\sigma v \rangle_\textrm{in}}\frac{(m_\textrm{n} + m_{\mathrm{i}})}{m_{\mathrm{i}} m_\textrm{n} n_{\mathrm{i}}} \\ \bigg(\frac{|\hat{B} \cdot \nabla P_{\textrm{c,p},j}|}{P_{\textrm{c,p},j}}n_{\textrm{1,p},j} + \frac{|\hat{B} \cdot \nabla P_{\textrm{c,e},j}|}{P_{\textrm{c,e},j}}n_{\textrm{1,e},j}\bigg)
\end{multline}
where the subscripts p and e represent the contributions made by the protons and electrons respectively. The energy density of electrons is much lower than that of protons, so the contribution of protons dominates.

We also consider NLL damping, which occurs when the thermal ions have a resonance with the beat wave formed by the interaction of two resonant Alfvèn waves. In this case, the damping rate is given by \cite{kulsrud_plasma_2005} to be
\begin{equation}    \Gamma_{\textrm{damp,nll}} = 0.3\Omega\frac{v_{\textrm{t},i}}{c}\bigg(\frac{\delta B}{B}\bigg)^2 \approx 0.3 \frac{8}{\pi}\frac{v_{\textrm{t},i} v_\textrm{p}^2}{c}\sigma_\parallel\,,
\label{eq:damp_nll}
\end{equation}
where $\Omega$ is the relativistic cyclotron frequency, $v_{\textrm{t},i}$ is the ion thermal velocity, and $\delta B/B$ is the magnetic field fluctuation. The above employs the quasilinear theory relation  $\nu_s = (\pi/8) \Omega (\delta B/B)^2$ for the scattering rate $\nu_s = v_p^2 \sigma_\parallel$.

In the NLL case, we set $2 \Gamma_{\textrm{stream}}(p_1) = \Gamma_{\textrm{damp,nll}}$, considering only proton contributions to wave growth.\footnote{\autoref{eq:growth} and \autoref{eq:damp_in} represent linear growth and damping rates of the wave amplitude. In this case, the growth rate is $2 \Gamma_{\textrm{stream}}(p_1) = \Gamma_{\textrm{damp,nll}}$ due to non-linear effects, see Chapter 11 in \cite{kulsrud_plasma_2005}.} Solving for $\sigma_\parallel$ then gives 
\begin{equation}
\sigma_{\parallel\textrm{,nll}}(p_{1,j}) = \frac{\pi}{8}\sqrt{\frac{|\hat{B} \cdot \nabla P_{\textrm{c},j}|}{v_{\textrm{A}, i}P_{\textrm{c},j}} \frac{\Omega_0 c}{0.3 v_{\textrm{t},i} v_\textrm{p}^2}\frac{m}{m_{\mathrm{i}}}\frac{n_1}{n_{\mathrm{i}}}}\,.
\end{equation}
Now when including both protons and electrons we have a quadratic expression in $\sigma_\parallel$, so
\begin{multline}
\sigma_{\parallel\textrm{,nll,tot},j} = \frac{\pi}{8}\sqrt{ \frac{\Omega_0 mc}{0.3 v_{\textrm{t},i} v_{\textrm{A}, i} v_\textrm{p}^2}\frac{1}{m_{\mathrm{i}} n_{\mathrm{i}}}} \\ \times \sqrt{\bigg(\frac{|\hat{B} \cdot \nabla P_{\textrm{c,p},j}|}{P_{\textrm{c,p},j}}n_{1, \textrm{p}, j} + \frac{|\hat{B} \cdot \nabla P_{\textrm{c,e},j}|}{P_{\textrm{c,e},j}}n_{1, \textrm{e}, j}\bigg)}\,.
\end{multline}

We solve for both $\sigma_{\parallel\textrm{,in,tot},j}$ and $\sigma_{\parallel\textrm{,nll,tot},j}$, then take the minimum of the two values (which corresponds to the stronger damping rate). This is a valid approximation, as in almost all regimes either IN or NLL damping will dominate. 

The parallel scattering coefficient,  $\sigma_\parallel$, determines transport aligned with the local magnetic field. We also consider a perpendicular scattering coefficient, $\sigma_\perp$, which represents scattering due to unresolved fluctuations in the mean magnetic field. Here, we consider two cases, one in which $\sigma_\perp = 10 \times \sigma_\parallel$  \citep[see Section~4.3 in][]{armillotta_cosmic-ray_2021} and another case with very little to no perpendicular diffusion with $\sigma_\perp \gg \sigma_\parallel$. For the $\sigma_\perp \gg \sigma_\parallel$ case, we set $\sigma_\perp$ to a very large, fixed value (1 cm$^{-2}$ s).
\cite{armillotta_cosmic-ray_2021} do not find a significant difference between these two cases. However, they considered only 1 GeV protons where the value of $\sigma_\perp = 10 \times \sigma_\parallel$ is relatively large ($\langle \sigma_\perp \rangle \lesssim 10^{-29}$~s~cm$^{-2}$ in the cold/warm ISM and $\gtrsim 10^{-27}$~s~cm$^{-2}$ in the warm/hot ionized gas).
When we consider higher energy CRs, the value of $\sigma_\parallel$ is greatly reduced ($\sigma \propto n_1$ or $n_1^{1/2}$), which in turn reduces the value of $\sigma_\perp$. In this case, we do see a difference in the final CR distribution depending on the definition of $\sigma_\perp$, as discussed in \autoref{sec:sigma_perp}.

\subsection{Ionization Fraction}
\label{sec:ion_frac}

In \autoref{sec:scattering} and \autoref{sec:losses} we use the electron, ion, and neutral densities of the gas to determine the value of the scattering coefficient and magnitude of CR energy losses. The electron density is defined as $n_e = x_e n_H$ where $x_e$ is the electron fraction. We find this value as in \cite{armillotta_cosmic-ray_2021}. For gas with $T > 2 \times 10^4$ K, the value of $x_e$ is interpolated from the values listed in \cite{sutherland_cooling_1993}, which is computed under the condition of collisional ionization equilibrium. 

For cooler gas, with $T \leq 2 \times 10^4$ K, the electron fraction is instead computed by balancing cosmic ray ionization with recombination for hydrogen, also allowing for low ionization potential metals to be ionized by far-UV \citep{draine_physics_2011}. The free electron abundance is
\begin{equation}
    x_e = x_{\mathrm{M}} + \frac{\sqrt{(\beta + \chi + x_{\mathrm{M}})^2 + 4\beta} - (\beta + \chi + x_{\mathrm{M}})}{2}\,.
    \label{eq:xe}
\end{equation}
Here $x_{\mathrm{M}} = 1.68 \times 10^{-4}$ is the adopted contribution from metals with ionization potential less than that of hydrogen (13.6 eV). The value of $\beta$ is $\zeta_H/(\alpha_{rr}n_H)$ with $\zeta_H$ the CR ionization rate per hydrogen atom and $\alpha_{rr} = 1.42 \times 10^{-12} \textrm{ cm}^{3} \textrm{ s}^{-1}$ the rate coefficient for radiative recombination. $\chi$ is the ratio $\alpha_{\rm gr}/\alpha_{\rm rr}$ where $\alpha_{\rm gr} = 2.83 \times 10^{-14} \textrm{ cm}^{3} \textrm{ s}^{-1}$ is the grain assisted recombination rate coefficient.

The CR ionization rate, $\zeta_H$, includes ionization due to CR nuclei and secondary electrons produced by primary ionization events. This value can be approximated as $\zeta_H \approx 1.5 \zeta_c$, with $\zeta_c$ the ionization rate due to primary events \citep[e.g.][]{padovani_impact_2020}. We compute $\zeta_c$ as
\begin{equation}
    \zeta_\mathrm{c} = \zeta^\mathrm{obs} \frac{e_\mathrm{c,0}}{e_\mathrm{c,0}^\mathrm{obs}}\,,
\end{equation}
where $\zeta^\mathrm{obs} = 2\times 10^{-16} \textrm{ s}^{-1}$ is the average CR ionization rate measured in the local ISM \citep[e.g.][]{indriolo_h3_2007}. $e_\mathrm{c,0}$ and $e_\mathrm{c,0}^\mathrm{obs}$ are, respectively,  the simulated and observed energy densities of CRs within the lowest-momentum bin, centered in $p_0 = 2$~GeV/c. To determine $e_\mathrm{c,0}^\mathrm{obs}$, we use the approximation in \autoref{eq:endens}, with $p_\mathrm{bin} = p_0 $, and $f(p_\mathrm{bin})$
the observed distribution function for protons \citep[obtained from Equation~1 in][]{padovani_cosmic-ray_2018} evaluated at $p_0$.

We define the ion number density in an analogous way to $n_e$ as $n_{\mathrm{i}} = x_{\mathrm{i}} n_H$ where $x_{\mathrm{i}}$ is the ion fraction. For $T \leq 2 \times 10^4$ K, \autoref{eq:xe} can also be used to estimate the value of $x_{\mathrm{i}}$ as it considers only singly ionized species, so that $x_e = x_{\mathrm{i}}$. For $T > 2 \times 10^4$ K, we approximate the ion fraction directly from the electron fraction taking $x_{\mathrm{i}} = \textrm{min}(x_e, 1.099)$. The maximum value 1.099 comes from the simplifying assumption that the number of free electrons is dominated by the ionization of hydrogen and helium, and that the hydrogen to helium ratio is approximately 10:1. From the ion fraction, we define the ion density as $\rho_{\mathrm{i}} = \mu_{\mathrm{i}} m_p n_{\mathrm{i}}$ where $\mu_{\mathrm{i}}$ is the ion mean molecular weight, defined as
\begin{equation}
\mu_{\mathrm{i}} = \frac{x_{\textrm{H}^+} + 4(x_{\textrm{He}^+} + x_{\textrm{He}^{++}}) + 12x_{\textrm{M}}}{x_{\mathrm{i}}}\,,
\end{equation}
where we approximate the mean atomic mass number of metals to be 12. To find $x_{\textrm{H}^+}$ and $x_{\textrm{He}^+} + x_{\textrm{He}^{++}}$, we must assume that all hydrogen is ionized before any helium. Therefore, if $x_{\mathrm{i}} - x_{\mathrm{M}} < 1$ then $x_{\textrm{H}^+} = x_{\mathrm{i}} - x_{\mathrm{M}}$. If all hydrogen is ionized, $x_{\mathrm{i}} - x_{\mathrm{M}} > 1$, and $x_{\textrm{H}^+} = 1$. In this case, the fraction of ionized helium is defined as,  $x_{\textrm{He}^+} + x_{\textrm{He}^{++}} = x_{\mathrm{i}} - x_{\mathrm{M}} - 1$. 

Once we have the ion density, the neutral density is then defined as $n_{\mathrm{n}} m_{\mathrm{n}} = \rho - \rho_{\mathrm{i}}$. The magnitude of some CR losses (discussed in \autoref{sec:losses}) depends on the fraction of neutral hydrogen rather than the total neutral density. Therefore, we define
\begin{equation}
    x_{\mathrm{n}} = \frac{n_{\rm{HI}} + 2 n_{\rm{H2}}}{n_H}\,.
\end{equation}
We approximate this value as we do for $\mu_{\mathrm{i}}$ by assuming that if $x_{\mathrm{i}} - x_{\mathrm{M}} > 1$, then all hydrogen is ionized and $x_{\mathrm{n}} = 0$. Otherwise, $x_{H^+} = x_{\mathrm{i}} - x_{\mathrm{M}}$ and $x_{\mathrm{n}} = 1 - x_{H^+}$.

\subsection{Energetic Losses}
\label{sec:losses}

Both CR protons and electrons lose energy to the surrounding ISM, for example through collisions with ambient atoms or through interactions with magnetic or radiation fields. We include losses as in \cite{armillotta_cosmic-ray_2021} with updates to include analytic expressions for different loss mechanisms.

Individual CRs lose energy at a rate given by
\begin{equation}
    \left[\frac{dE}{dt}\right]_j = -v_{c,j} L(E_{j}) n_H \equiv -\Lambda(E_{j}) E_{j} n_H\,,
    \label{eq:lossrate}
\end{equation}
with $E_{j}$ the relativistic energy of the particle in bin $j$ and $v_{c, j}$ the CR velocity (\autoref{eq:v_c}). $L(E_{j})$ is the energy loss function described in \cite{padovani_impact_2020}. 

Since CRs in the same momentum bin share the same energy, the energy density lost from CRs in a given bin is calculated as
\begin{equation}
    \dot{e}_{\textrm{c}, j,\rm loss} = - \Lambda(E_{j}) n_H e_{\textrm{c}, j}\,.
    \label{eq:energy_loss}
\end{equation}
This total loss rate is applied to the right-hand side of \autoref{eq:energy}. The CR fluxes are updated correspondingly by applying the following term to the right-hand side of \autoref{eq:flux}
\begin{equation}
    \dot{F}_{\textrm{c}, j, \rm loss} = - \Lambda(E_{j}) n_H F_{\textrm{c}, j} / v_{\textrm{c}, j}^2\,.
    \label{eq:flux_loss}
\end{equation}
The loss function $\Lambda$ includes the contribution from all sources of energy losses. A summary of these follows, with the exact expressions provided in \autoref{sec:app_p_losses} and \autoref{sec:app_e_losses} for protons and electrons, respectively.\footnote{\autoref{eq:energy_loss} and \autoref{eq:flux_loss} are equivalent to Equation 11 in \cite{armillotta_cosmic-ray_2021} but we introduce a new, clearer notation.}

At low kinetic energies, energy losses of CR protons are dominated by either the ionization of neutral hydrogen or Coulomb interactions with free electrons. These two sources are relevant in the warm/cold, primarily neutral ISM or the hot, well-ionized medium respectively. At higher kinetic energies, greater than $\sim$1 GeV, losses are instead dominated by pion production due to collisions with atoms in the surrounding ISM. The total value of $\Lambda$ for protons is then given by
\begin{equation}
    \Lambda_{\textrm{p}, j} = \Lambda_{\textrm{pion}, j} + \Lambda_{\textrm{ion}, j} + \Lambda_{\textrm{Coulomb}, j}.
\end{equation}

Similarly to protons, low energy CREs lose their energy primarily through ionization of neutral hydrogen and Coulomb interactions. In addition to these losses, CREs are also subject to energy losses due to bremsstrahlung interactions. At higher energies, synchrotron and IC losses dominate.  The total $\Lambda$ for electrons is given by
\begin{equation}
    \Lambda_{\textrm{e}, j} = \Lambda_{\textrm{synch}, j} + \Lambda_{\textrm{IC}, j} + \Lambda_{\textrm{ion}, j} + \Lambda_{\textrm{Coulomb}, j} + \Lambda_{\textrm{brem}, j}.
\end{equation}

As discussed in \autoref{sec:transport}, the loss of energy by CRs in a given energy bin should result in energy gain in a lower energy bin. However, we do not include transfer of energy between CR energy bins due to these losses. The CR distribution follows a steep power law slope. Therefore, the number of CRs that would be transferred from a higher to a lower energy bin is not significant compared to the number of CRs injected by SNe.

\subsection{CRE Spectrum and Synchrotron Emission}
\label{sec:synch}

We use the final values of $e_{c, j}$ to estimate the CRE spectrum, $j_e$. The latter is related to the CRE distribution function as $j_\mathrm{e}(E) = v_\mathrm{e} f_\mathrm{e}(p) p^2 dp/dE$. If we substitute this expression in \autoref{eq:endens}, we obtain
\begin{equation}
    j_e(E_{j}) \approx \frac{e_{\textrm{c}, j} v_{\textrm{e}, j}}{4\pi E_{j}^2 d\textrm{ln}E}\,,
\label{eq:spectrum}    
\end{equation}
where $d\textrm{ln}E$ is the width of the bin in log space. In the relativistic limit $d\textrm{ln}E \approx (p^2 c^2 / E^2) d\textrm{ln}p \approx d\textrm{ln}p$. 
The value of $j_e$ is in units of the number per unit energy, time, area, and solid angle. To estimate the CRE spectrum at all energies we interpolate between the momentum bins, approximating $j_e$ as a broken power law.

We use the simulated CRE spectrum to produce synthetic synchrotron emission using the method described in \cite{padovani_spectral_2021}, which we summarize here \citep[see also][]{ponnada_synchrotron_2023}. The specific emissivity (i.e. per unit volume per unit time per unit solid angle per unit frequency) from each cell is given by
\begin{equation}
    \epsilon_{\nu,\parallel} = \int_{m_e c^2}^\infty \frac{j_e(E)}{v_e(E)}P_{\nu,\parallel}(E)dE
    \label{eq:eps_par}
\end{equation}
\begin{equation}
    \epsilon_{\nu,\perp} = \int_{m_e c^2}^\infty \frac{j_e(E)}{v_e(E)}P_{\nu,\perp}(E)dE
    \label{eq:eps_perp}
\end{equation}
where the parallel and perpendicular components are relative to the line of sight. Here, the emissivity is found by integrating over all CRE energies, but we only simulate CREs with energies between 2-101 GeV. We extrapolate to a range of 1-$10^3$ GeV by assuming a constant power law slope at energies above and below our simulated bins. This approximation is consistent with direct observations \citep[e.g.][]{padovani_spectral_2021}. However, we are limited to modeling synchrotron frequencies in which the emissivity is dominated by CREs in our simulated energy range. 

The synchrotron power per unit frequency emitted at frequency $\nu$ is given by 
\begin{equation}
    P_{\nu,\parallel} = \frac{\sqrt{3}e^3}{2 m_e c^2} B_\perp (F(x) - G(x))
\end{equation}
\begin{equation}
    P_{\nu,\perp} = \frac{\sqrt{3}e^3}{2 m_e c^2} B_\perp (F(x) + G(x))
\end{equation}
where $B_\perp$ is the magnitude of the magnetic field perpendicular to the line of sight. F(x) and G(x) are the synchrotron functions given by 
\begin{equation}
    F(x) = x \int_x^\infty K_{5/3}(\xi)d\xi
\end{equation}
\begin{equation}
    G(x) = x K_{2/3}(x)
\end{equation}
in which $K_{5/3}$ and $K_{2/3}$ are modified Bessel functions of order 5/3 or 2/3 and $x = \nu/\nu_c$ with $\nu_c$ the critical frequency,
\begin{equation}
    \nu_c = \frac{3 e B_\perp}{4 \pi m_e c}\bigg(\frac{E}{m_e c^2}\bigg)^2.
\end{equation}

The values of F(x) and G(x) are taken from tables provided by Padovani et al.\ (priv. comm.) \citep[also tabulated in][]{shu_radiation}. The emissivities can be integrated along the line of sight to find the specific intensities $I_{\nu,\parallel}$ and $I_{\nu,\perp}$with the total specific intensity $I_\nu = I_{\nu,\parallel} + I_{\nu,\perp}$.

If the CR distribution were described by a single power law in energy with $j_e \propto E^{s}$ (where $j_e \propto f E^2$ in the relativistic regime) then the synchrotron intensity will also be a power law with $I_\nu \propto \nu^{-\alpha}$ where $\alpha = -({s+1})/{2}$ (see e.g. \cite{shu_radiation} equation 19.31). Note that with $f(p) \propto p^{-\gamma}$ (i.e. $\gamma=2-s$ in the relativistic regime) this is equivalent to $\alpha = (\gamma - 3)/2$.

\section{Results}
\label{sec:results}

\subsection{CR Distribution}
\label{sec:results_dist}

\begin{figure*}
    \centering
    \includegraphics[scale = 0.58]{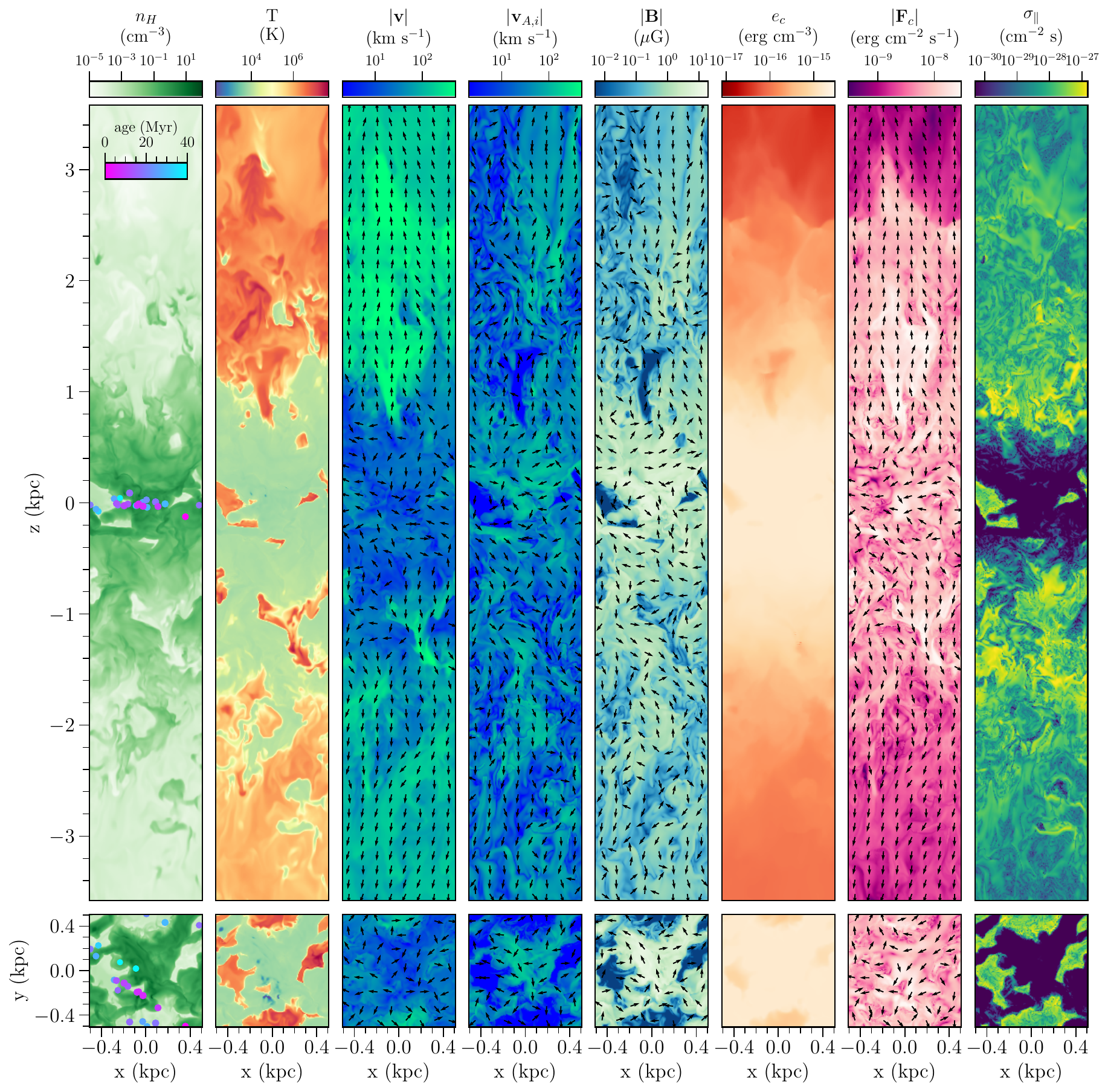}
    \caption{Vertical (through $y=0$) and horizontal (through $z=0$) slices of MHD and CR variables from the snapshot at $t = 214$~Myr. From left to right, the first five slices show number density of hydrogen ($n_H$), temperature (T), magnitude of the gas velocity ($|\mathbf{v}|$), magnitude of the ion Alfvén velocity ($|\mathbf{v}_\mathrm{A,i}|$), and magnitude of the magnetic field ($|\mathbf{B}|$). The final three slices show energy density ($e_c$), flux magnitude ($|\mathbf{F}_c|$), and scattering coefficient ($\sigma_\parallel$) of CREs with momentum $p = 2$~GeV/c. The arrows overlaid on the slices of gas velocity, Alfvén speed, magnetic field, and flux magnitude indicate the projected directions of the gas velocity, CR streaming velocity, magnetic field, and CR flux respectively.}
    \label{fig:snapshot}
\end{figure*}

In \autoref{fig:snapshot}, we present vertical and horizontal slices (through $y=0$ and $z=0$ respectively) of relevant MHD and CR variables. These quantities are taken from the TIGRESS snapshot at $t = 214$~Myr following both post-processing and subsequent MHD relaxation. The first column shows the hydrogen number density, $n_H$, along with the young star clusters colored by their age. The second panel shows the gas temperature, $T$. From these two panels, we can see the transition from the disk midplane region, mostly composed of warm and cold moderate-density gas ($T\lesssim 10^4$ K, $n_H\sim 0.1-100\ {\rm cm}^{-3}$), to the extraplanar region, where most of the volume is occupied by  hot, diffuse gas ($T\gtrsim 10^6$ K, $n_H\lesssim 10^{-3}\ {\rm cm}^{-3}$). The star clusters, which act as CR sources, are within $|z|\lesssim200$ pc.

The next three panels represent the magnitudes of the gas flow speed $v$, ion Alfvén speed $v_\mathrm{A,i}$ (equal to the magnitude of the CR streaming velocity $\mathbf{v}_\mathrm{s}$), and magnetic field strength $B$, respectively. The arrows overlaid on each of these panels indicate the projected direction of the CR streaming velocity, gas velocity, and magnetic field, respectively. Compared to $v$, the magnitude of $v_\mathrm{A,i}$ is larger in cool, high-density, primarily neutral gas, and much smaller in hot, low-density, ionized gas. Therefore, while CR advection dominates in the hot phase, streaming can be important in the cooler gas \citep[see also][]{armillotta_cosmic-ray_2021, armillotta_cosmic-ray_2024}. In general, $v_\mathrm{A,i}$ is more turbulent than $v$. Above $z\sim1$ kpc, the gas velocity is predominantly oriented outward from the simulation box, due to the presence of strong outflows. Although there is some organization of $v_\mathrm{A,i}$ pointing out of the simulation box at large $z$, the alignment is not nearly as clear as in $v$. 

In the last three panels, we show the energy density $e_\mathrm{c}$, flux $F_\mathrm{c}$, and parallel scattering coefficient $\sigma_\parallel$ of the lowest energy CREs ($p =  2$~GeV/c) for the case where $\sigma_\perp = 10\times\sigma_\parallel$. As with the velocity and magnetic field, the $F_c$ panel has the projected vectors overlaid on the magnitude. The CR energy density distribution is very smooth compared to the gas density distribution, highlighting the importance of streaming and diffusive transport in addition to advection. Near the midplane, the CR flux is somewhat turbulent. Above $z\sim1$ kpc, however, ${\bf F}_c$ mostly aligns with the velocity streamlines, confirming that advection is the dominant propagation mechanism in the hot wind. In the denser neutral gas near the midplane, $\sigma_\parallel$ is small, meaning that diffusion is significant. In contrast, $\sigma_\parallel$ is relatively large in the diffuse, hot, ionized gas. 

\subsection{Choice of perpendicular scattering coefficient}
\label{sec:sigma_perp}

We model CR transport with two choices of the perpendicular scattering coefficient, $\sigma_\perp$: $\sigma_\perp = 10 \times \sigma_\parallel$ or $\sigma_\perp \gg \sigma_\parallel$ (see \autoref{sec:scattering}). In previous work considering only 1 GeV CRs, the choice of $\sigma_\perp$ did not lead to significant differences in the final CR distribution \citep{armillotta_cosmic-ray_2021}. When we consider higher energy CRs, however, we do see changes in the resulting CR distribution depending on the choice of scattering coefficient.

\begin{figure*}
    \centering
    \includegraphics[scale = 0.55]{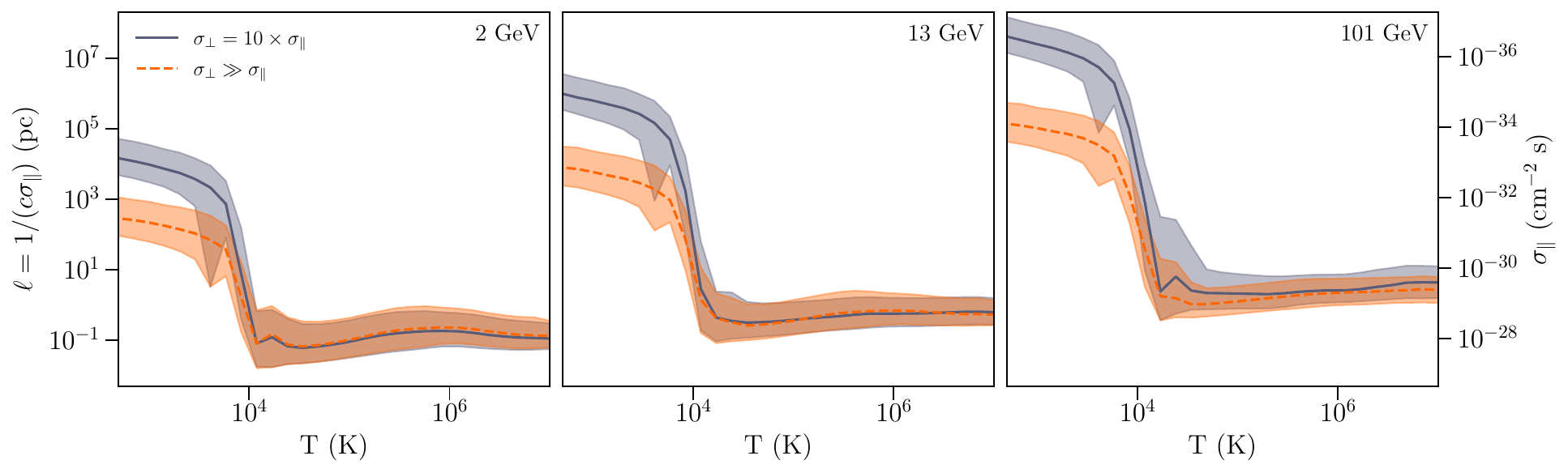}
    \caption{CR mean free path, $\ell = 1/(c \sigma_\parallel)$ and $\sigma_\parallel$ as a function of gas temperature, T. The lines represent the median value across all snapshots, while the shaded regions represent the 16th-84th percentiles. Each panel represents a different CR momentum bin, $pc = 2$, 13, and 101 GeV. In each bin, we include the distribution of $\ell$ and $\sigma_\parallel$ for both $\sigma_\perp = 10 \times \sigma_\parallel$ (gray, solid lines) and $\sigma_\perp \gg \sigma_\parallel$ (orange, dashed lines).}
    \label{fig:mfp_sigma}
\end{figure*}

In \autoref{fig:mfp_sigma}, we present the value of the CR mean free path, $\ell = 1/(c \sigma_\parallel)$, along with $\sigma_\parallel$ as a function of gas temperature, $T$. Each of the three panels represents one of our momentum bins at $E \approx pc = 2$, 13, and 101 GeV, corresponding to the lowest, middle, and highest momentum bins that we simulate. Within each panel, we compare the value of $\ell$ for the two definitions of $\sigma_\perp$. At low temperatures, IN is the dominant damping mechanism while at higher temperatures, $T \gtrsim 10^4$ K, NLL dominates. The transition between these two regimes is evident through a drop by more than four orders of magnitude of $\ell$.

For $T \gtrsim 10^4$ K, $\ell$ is the same for either choice of $\sigma_\perp$. At these temperatures, $\ell$ is short so the CRs are well coupled to the gas and advection is the dominant mechanism for CR transport. If T $\lesssim 10^4$ K, however, diffusion is more important than advection. In this regime, $\ell$ is consistently shorter when $\sigma_\perp \gg \sigma_\parallel$. With this choice of $\sigma_\perp$, the CRs are less diffusive, and larger CR energy gradients form. This increases $\nabla P_{\rm c}$ thereby increasing $\sigma_\parallel$ and further reducing the diffusivity of the CRs. 

\begin{figure*}
    \centering
    \includegraphics[scale = 0.7]{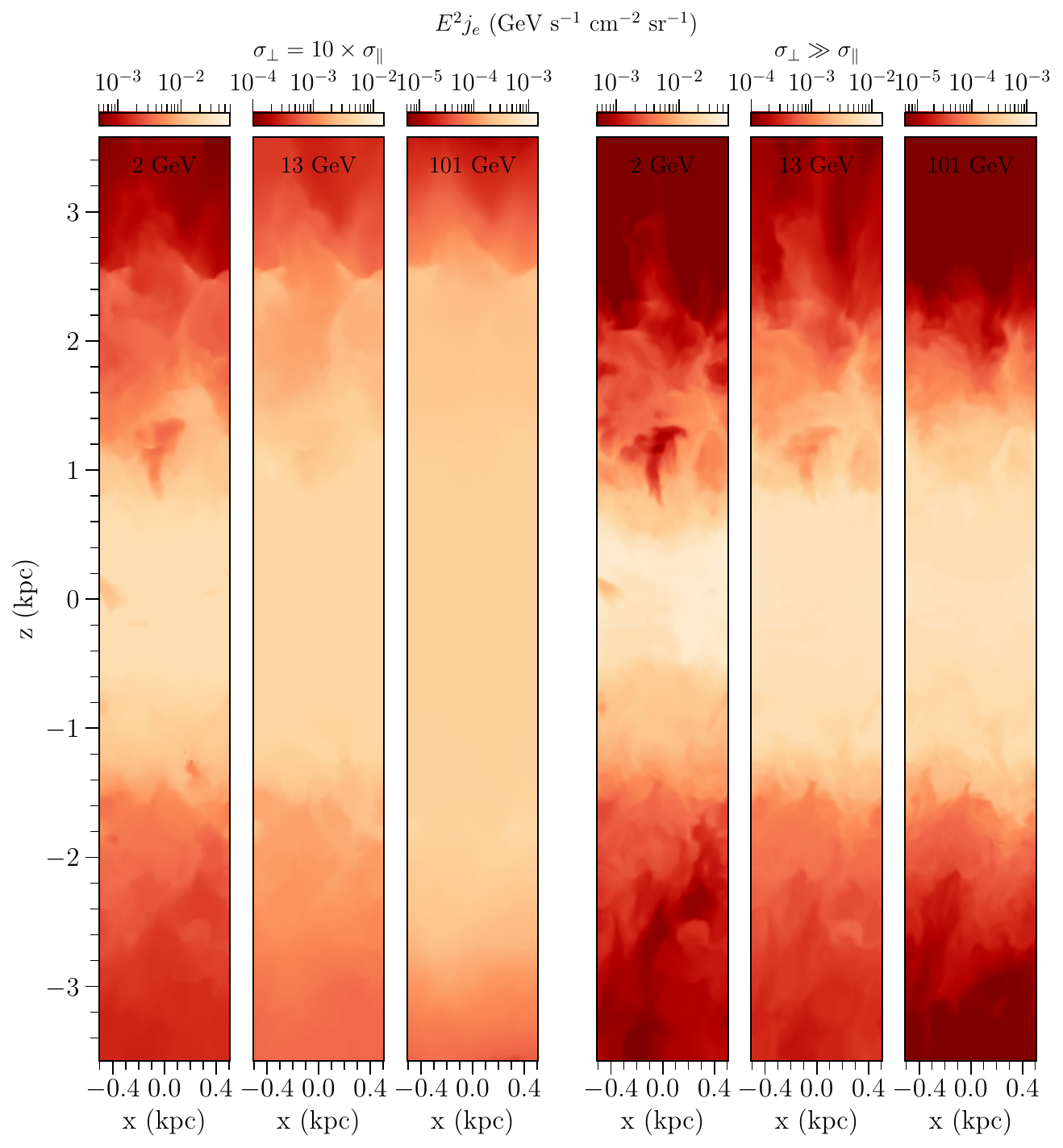}
    \caption{Vertical slices (through $y = 0$) of the CRE energy spectrum ($E^2 j_e$) in the bins centered at $pc = 2$, 13, and 101 GeV taken from the snapshot at $t=214$~Myr. For each bin, we include the results for both $\sigma_\perp = 10 \times \sigma_\parallel$ (left three panels) and $\sigma_\perp \gg \sigma_\parallel$ (right three panels).}
    \label{fig:sigma_perp}
\end{figure*}

The difference in the rate of diffusion between the two choices of $\sigma_\perp$ is evident in the spatial distribution of the CRE energy spectrum, $E^2 j_e \propto e_c$ (see \autoref{eq:spectrum}), which we present in \autoref{fig:sigma_perp}. We compare vertical slices through the simulation box at $y = 0$ at the same energy values as in \autoref{fig:mfp_sigma} ($pc = 2$, 13, and 101 GeV). All slices are taken from the snapshot at $t = 214$~Myr. 

The distribution of the lowest energy CREs is similar in each of the two cases (as in \citealt{armillotta_cosmic-ray_2021}). The values of $\sigma_\parallel$ and $\sigma_\perp$ are large enough in both definitions to limit significant diffusion. As energy increases, however, $\sigma_\parallel$ decreases and the value of $\sigma_\perp = 10 \times \sigma_\parallel$ becomes much smaller compared to the large, fixed value we set for $\sigma_\perp \gg \sigma_\parallel$. The difference in $\sigma_\parallel$ between the two cases also becomes larger (\autoref{fig:mfp_sigma}). Therefore, there is significantly more diffusion if $\sigma_\perp = 10 \times \sigma_\parallel$ case compared to $\sigma_\perp \gg \sigma_\parallel$. In the highest momentum bin, the distribution of CREs when $\sigma_\perp = 10 \times \sigma_\parallel$ is practically uniform, whereas the $\sigma_\perp \gg \sigma_\parallel$ case still shows significant structure with greater $j_\mathrm{e}$ near $z = 0$. This result is true across all the snapshots. 

\begin{figure}
    \centering
	\includegraphics[scale = 0.45]{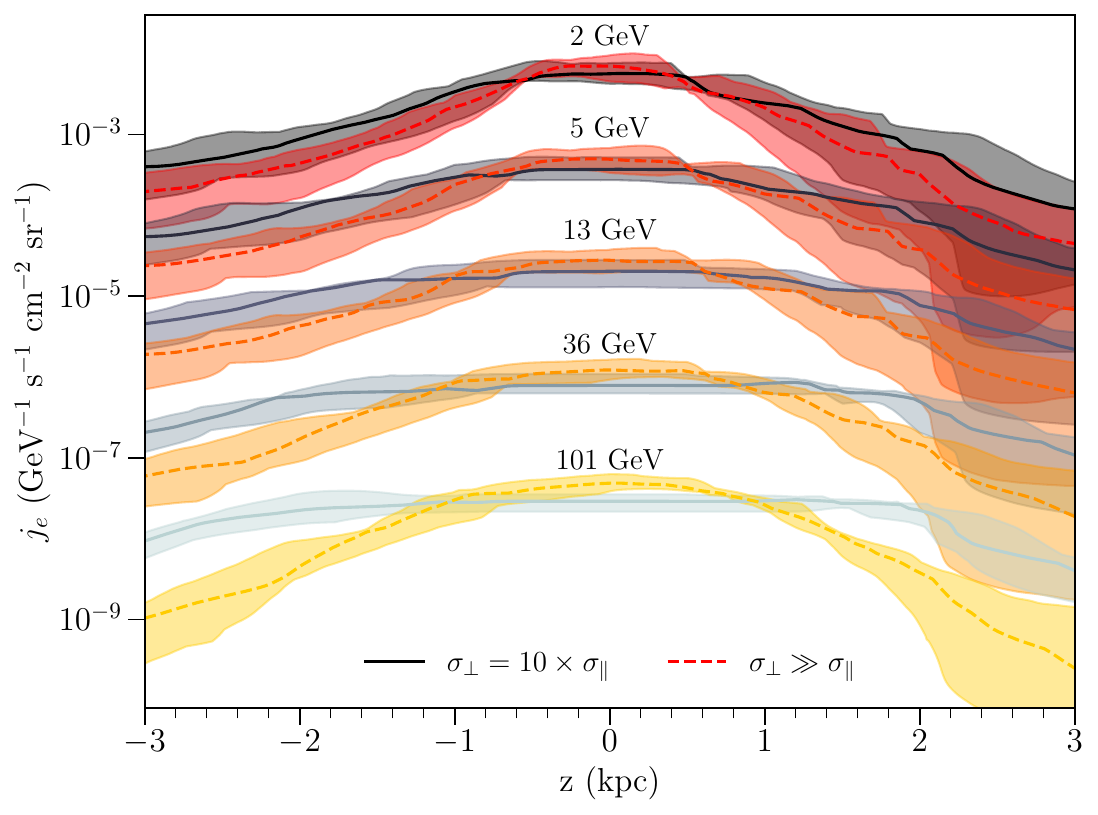}
    \caption{Horizontally and temporally averaged vertical profiles of the CRE spectrum, $j_e$, in each momentum bin. The shaded area covers the 16th and 84th percentiles from temporal variations while the central line represents the median value. The solid lines with a gray color scale represent the simulations with $\sigma_\perp = 10 \times \sigma_\parallel$. The dashed lines with a red-yellow color scale represent simulations with $\sigma_\perp \gg \sigma_\parallel$.}
    \label{fig:zprof}
\end{figure}

In \autoref{fig:zprof}, we show horizontally averaged profiles of the CRE spectral flux $j_\mathrm{e}$ at different momenta as a function of $z$ for both values of $\sigma_\perp$. The profiles are computed as follows: within each snapshot, we find the horizontally averaged value of $j_e$ at each $z$, denoted as $\langle j_e \rangle$; we then determine the median and 16th-84th percentile values of $\langle j_e \rangle$ across all snapshots. For all momenta, the profiles show a smaller scale height when $\sigma_\perp \gg \sigma_\parallel$ than when $\sigma_\perp = 10 \times \sigma_\parallel$, with the difference in scale height increasing with CRE momentum. In the highest momentum bin, the CRE spectrum is practically uniform across $z$ when $\sigma_\perp = 10 \times \sigma_\parallel$, owing to the high diffusion. Near the midplane, however, the magnitude of $j_e$ is very similar for both $\sigma_\perp$ conditions in all bins. Therefore, we conclude that the results derived from the CR distribution near the midplane will be consistent regardless of the condition on $\sigma_\perp$.

\subsection{Energy Losses}
\label{sec:results_loss}

\begin{figure}
    \centering
	\includegraphics[scale = 0.57]{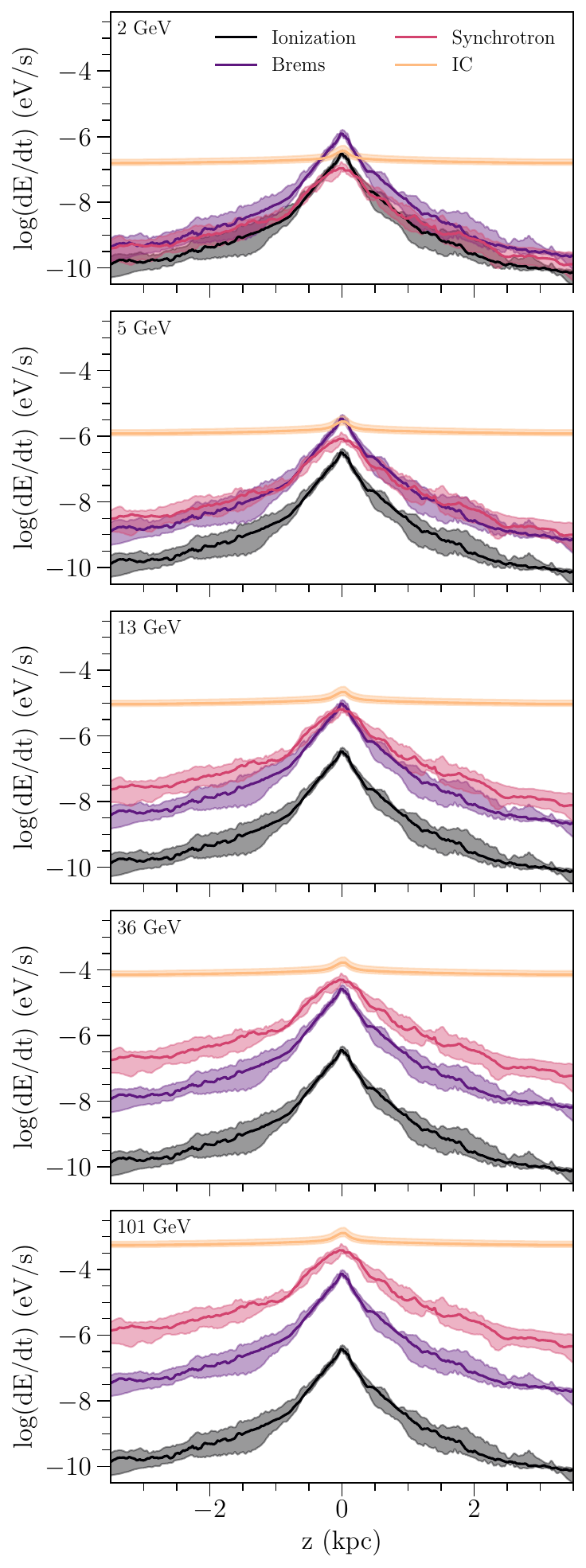}
    \caption{Horizontally and temporally averaged vertical profiles of CRE energy loss rates $dE/dt$. Different colors represent different loss mechanisms: ionization (black), bremsstrahlung (purple), synchrotron (magenta), and IC (light orange). Each panel shows results for CREs within a different momentum bin. At each $z$, the solid lines represent the median value of the horizontally averaged $\langle {dE}/{dt} \rangle$ evaluated across all snapshot times. The shaded regions cover the 16th-84th percentiles.}
    \label{fig:losses}
\end{figure}

In CR electrons, unlike protons, the rate of energy loss is extremely important for determining the steady state CR spectrum. We show the horizontally and temporally averaged values of the CRE energy loss rate, $\left[dE/dt\right]_j$ (\autoref{eq:lossrate}), as a function of height for each CRE momentum bin $j$ in \autoref{fig:losses}. Different lines represent different loss mechanisms as described in \autoref{sec:losses} (and in more detail in \autoref{sec:app_e_losses}): ionization, bremsstrahlung, synchrotron, and IC. As in \autoref{fig:zprof}, we compute the vertical profiles by evaluating the median and 16th-84th percentile of the horizontally averaged $\langle dE/dt \rangle$ across all snapshots.

For the lowest momentum CREs, all loss rates are roughly comparable near the midplane. As CR momentum increases, however, the relative scaling of the loss mechanisms leads to IC dominating. Synchrotron losses have the same energy scaling as IC ($\propto \gamma^2$), but the magnetic energy density is generally lower than the photon energy density in the solar-neighborhood TIGRESS simulations, so the IC losses dominate overall. Above z $\approx$ 500 pc, the IC losses dominate at all energies. IC dominates losses at large $|z|$ because the photon energy density drops off much less steeply with $z$ than either the magnetic energy density (relevant for the synchrotron losses) or number density of the gas (important for the bremsstrahlung and ionization losses).

Using these loss rates, we define a loss timescale as 
\begin{equation}
    \label{eq:loss_time}
    t_{\textrm{loss}, j}(z) = \frac{E_j}{\left[\frac{dE}{dt}\right]_{j}(z)}\,,
\end{equation}
where the denominator includes all four loss mechanisms. This is an approximation for the time it would take a CRE at height $z$ to lose all of its energy. It does not account for the fact that $dE/dt$ would change as the CRE loses energy, or that the CRE can propagate in space. 

We compare this loss timescale to an estimate of the transport time defined as
\begin{equation}
    \label{eq:transport_time}
    t_{\textrm{transport}, j}(z) = H_{\rm gas }\frac{4}{3}\frac{\langle e_{\textrm{c},j}(z) \rangle}{\langle |F_{\textrm{c}, z, j}(z)| \rangle }\,,
\end{equation}
which represents the average time it takes for a CRE to traverse the scale height of the gas. The ratio $\frac{4}{3}\langle e_{\textrm{c},j}(z) \rangle/\langle |F_{\textrm{c}, z, j}(z)| \rangle$ is the inverse of the effective mean vertical propagation speed, $v_\mathrm{eff,z,j}$, with $\langle e_{\textrm{c},j}\rangle$ and $\langle |F_{\textrm{c}, z, j}(z)| \rangle$ the horizontally averaged CR energy density and vertical component of the CR flux respectively. $H_{\rm gas}$ is the scale height of the horizontally averaged gas density. If we wanted to estimate the total CR transport time out of the galaxy, we would need to multiply this expression by a factor of $H_{\rm c}/H_{\rm gas}$, where $H_{\rm c}$ is the CR scale height.

The transport time is most limited by downstream regions (at larger $|z|$) where $v_\mathrm{eff,z,j}$, due to the combination of advection, streaming, and diffusion, is smallest. \autoref{fig:snapshot} shows that the scattering coefficient $\sigma_\parallel$ is quite small in most of the midplane region (where the gas is mostly neutral), becoming large only at $|z|\gtrsim 0.5$ kpc. Because all CRs have to pass through the high-scattering region at $|z|\gtrsim 0.5$ kpc in order to escape from the box, this  limits the net vertical flux and hence $v_\mathrm{c,eff,z}$ in the ``upstream'' region near the midplane as well, even though the scattering rate is small there so that the CRs are highly diffusive. 

Thus, it is also useful to consider the diffusion time 
\begin{equation}
    \label{eq:transport_time_local}
    t_{\textrm{diff}, j}(z) = H_{\rm gas }\frac{4}{3}\frac{\langle e_{\textrm{c},j}(z) \rangle}{\langle |F_{\textrm{diff,}z, j}(z)| \rangle }\,,
\end{equation}
where $\langle |F_{\textrm{diff,}z, j}| \rangle$ is the horizontal average of the diffusive flux in the vertical direction, i.e. $|{\stackrel{\leftrightarrow}{\sigma_{j}}}^{-1} \cdot \hat{z}\partial P_c/\partial z|$. This is the time it would take a cosmic ray to traverse the scale length of the gas considering only diffusion.  Previously, \citet{armillotta_cosmic-ray_2024} (see Eq. 16 and Fig. 8 of that paper) showed that the diffusion speed for GeV CRs in the majority of the neutral gas near the midplane is $\sim 10^2  \;\mathrm{ km\ s^{-1}}$, which with $H_\mathrm{gas}= 300$ pc would imply a diffusion time of a few Myr.  Since higher energy CRs have even lower scattering rate, we would expect their diffusion timescales to be even lower.  

\begin{figure}
    \centering
	\includegraphics[scale = 0.57]{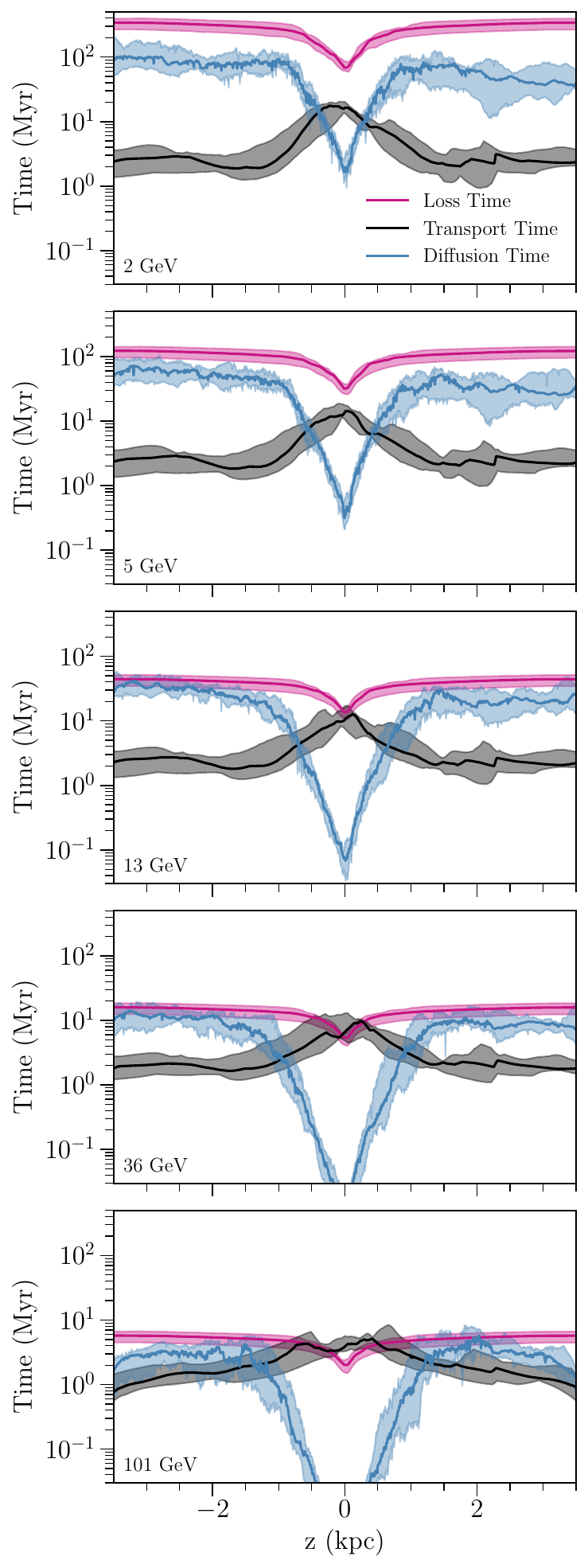}
    \caption{Horizontally and temporally averaged vertical profiles of the timescale for energy loss (magenta) in comparison to transport (black) and diffusion (blue). These timescales are defined in \autoref{eq:loss_time}, \autoref{eq:transport_time}, and \autoref{eq:transport_time_local}, respectively. Each panel presents results for CREs within a given momentum bin. At each $z$, the solid line represents the median value of the horizontally averaged $\langle t \rangle$ evaluated across all snapshots. The shaded regions show the 16-84th percentile values across all snapshots.}
    \label{fig:timescale}
\end{figure}

We show a comparison of the loss time with the transport and diffusion timescales for CRs in each momentum bin in \autoref{fig:timescale}. The loss timescale is minimized at the midplane where the loss rates are greatest (\autoref{fig:losses}). At larger $z$, this timescale is roughly constant with $z$ due to constant IC losses. As CRE momentum increases, energetic losses increase dramatically, so the timescale shortens correspondingly. For the highest momentum CRs, the loss timescale is only a few Myr at $z=0$.

Unlike the loss timescale, the transport time is maximized at $z = 0$ and decreases as $|z|$ increases  due to the increasing value of the advection speed of the gas (see \autoref{fig:snapshot}). As CR momentum increases, the transport time decreases slightly because, unlike the advection and streaming velocities which are momentum-independent, the diffusion velocity increases with CR momentum.  The decrease in transport time at higher energy means that even in the absence of losses, the actual CR spectrum would be steeper than the injection spectrum.  

For the lowest momentum CREs, the transport time is faster than the loss timescale at all $z$. At higher momentum, however, the loss timescale decreases such that it is comparable to the transport time near the midplane for the three highest momentum bins. We conclude that it is a combination of both the energy dependent losses and the reduction in transport time with increasing CRE energy that drives the steepening of the electron spectrum compared to the injected spectrum (see \autoref{sec:spectrum_result}).

Near the midplane, the CRE distribution is highly uniform, as seen in \autoref{fig:sigma_perp} and \autoref{fig:zprof}. This indicates that diffusion may be a dominant process in this region, and indeed \cite{armillotta_cosmic-ray_2024} showed quantitatively that for GeV protons, diffusion is the dominant transport mechanism in the warm/cold neutral medium found near the midplane. \autoref{fig:timescale} shows that the local diffusion time in the midplane region is significantly lower than the transport and loss timescales for the CREs in all momentum bins.  As discussed above, even though CRs easily diffuse along field lines in the neutral midplane region (leading to a short local transport time), the net flux out of the box is limited by the much higher scattering rate at large $|z|$ where gas is ionized. This means that CRs effectively transverse the midplane region many times before they are able to escape from the disk.

\subsection{CR Spectrum}
\label{sec:spectrum_result}

\begin{figure}
    \centering
        \includegraphics[scale = 0.55]{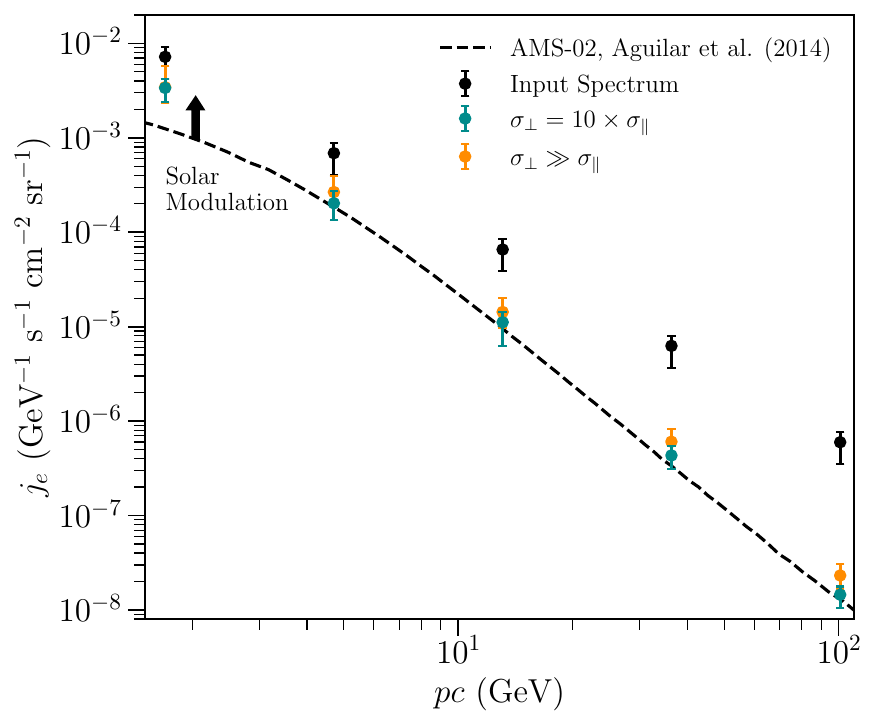}
    \caption{Spatially and temporally averaged CRE spectrum in the warm gas ($T < 3 \times 10^4$~K) within the disk region ($\vert z \vert < 300$~pc). The points represent the median value across space and time, while the error bars show the 16th-84th percentiles. The black points represent the input spectrum with a fixed power-law slope of $s = -2.3$. The blue and orange points represent the final CRE spectra for the two different conditions for $\sigma_\perp$. The normalization of the CRE spectrum has been reduced by a factor of two to more closely match the observed values from AMS shown in the black dashed line \citep{aguilar_electron_2014}.
    }
    \label{fig:spectrum}
\end{figure}

We compare the simulated CRE spectrum to the directly observed, solar neighborhood values in \autoref{fig:spectrum}. To best recreate typical conditions representative of the ISM we use only the warm gas ($T < 3\times10^4$ K) in the disk region ($|z| < 300$ pc). The limit on temperature removes the hot gas surrounding the injection regions \citep[e.g.][]{redfield_structure_2004}, but does not significantly change the results. We limit our comparison to the midplane region of the simulation both to mimic solar neighborhood conditions, and because our post-processing simulations do not necessarily reproduce accurate ISM conditions at larger $z$ (where CR pressure gradients may transfer momentum to the gas, accelerating it, see \autoref{sec:transport})

\autoref{fig:spectrum} includes the input CR spectrum as a function of $E\approx pc$, as well as the final, evolved spectra in both $\sigma_\perp$ cases. These values are averaged both spatially and across all snapshots (different snapshots have different SN rates and hence different CR injection rates and input spectra). Each point represents the median value, and the error bars are the 16th-84th percentile values. The spectra for $\sigma_\perp = 10  \times \sigma_\parallel$ and $\sigma_\perp \gg \sigma_\parallel$ are very similar, with the former slightly steeper than the latter. Indeed, as noted in \autoref{sec:sigma_perp}, the spectra are weakly dependent on the choice of $\sigma_\perp$ near the midplane. The final spectra are steeper than the initial spectrum due to both the energy-dependent transport and the energy-dependent losses that CREs experience as they propagate through the ISM (see \autoref{sec:results_loss}).

Along with the three simulated spectra, in \autoref{fig:spectrum}, we include direct observations of the CRE spectrum by AMS-02 \citep{aguilar_electron_2014}. To more closely match the observed values, we have rescaled the simulated spectra (including the input spectra) across all bins, reducing the magnitude by a factor of two. As discussed in \autoref{sec:injection}, the spectrum can be renormalized freely after the CR evolution to reflect a different initial normalization. 

Although the normalization of the simulated CRE spectra overestimate the observed value, the ratio to the simulated CR proton spectra is consistent with observations. We find the simulated proton spectra to also overestimate the observed values by a factor of two but reproduce the observed spectral slope well (Armillotta et al.\ 2025, accepted). Therefore, the adopted choice of a 2\% injection efficiency of electrons relative to protons is consistent with direct observations. The overestimation of the total CR energy density might be explained by an enhanced star formation rate (SFR) in the TIGRESS simulations ($\Sigma_\mathrm{SFR} \sim 5 \times 10^{-3}$~M$_\odot$~kpc$^{-2}$~yr$^{-1}$) compared to the mean solar neighborhood value over the last $100$~Myr ($2 \times 10^{-3} \lesssim \Sigma_\mathrm{SFR} \lesssim 5 \times 10^{-3}$~M$_\odot$~kpc$^{-2}$~yr$^{-1}$) \citep{zari_did_2023}, which is the period when the CRs we observe today were generated. Alternatively, if the SFR and SN rate in the Solar neighborhood are in agreement with the TIGRESS level, it would imply that the total energy injection per SN at $p_\mathrm{min}>1$ GeV/c would need to be reduced by a factor two relative to our assumptions, i.e. becoming $5\times 10^{49}$ erg for CR protons, and $10^{48}$ erg for CR electrons.  

The renormalized CRE spectrum matches the observations extremely well, especially in the four highest momentum bins. The lowest simulated momentum bin, however, has a larger magnitude than the observations. At energies below a few GeV, the AMS observations are affected by solar modulation such that the directly observed CR spectrum is reduced. This effect is also seen in other models of the CRE spectrum \citep[e.g.][]{padovani_cosmic-ray_2018, werhahn_cosmic_2021a}. 

\begin{figure}
    \centering
    \includegraphics[width=\linewidth]{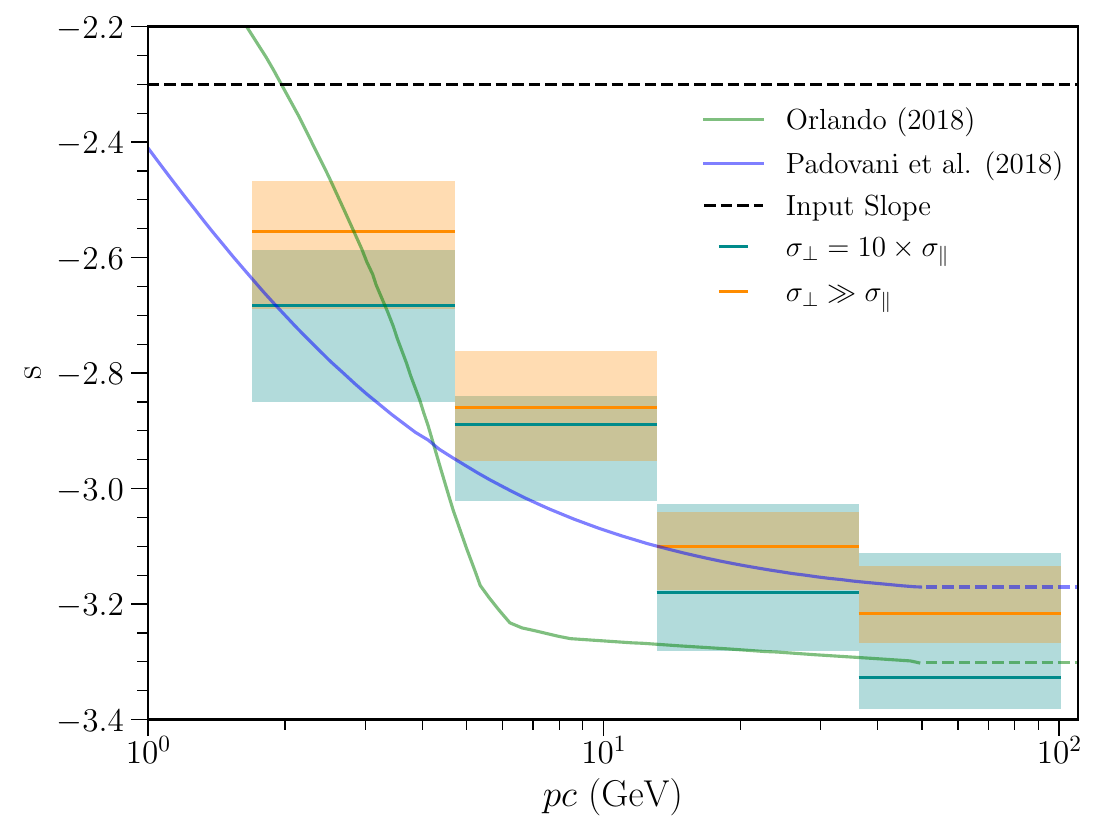}
    \caption{Power-law slope of the CRE spectrum (\autoref{fig:spectrum}) as a function of CR momentum. The slope is calculated across neighboring bins as in \autoref{eq:slope}. The solid horizontal lines represent the median value of the slope across all cells with $|z| < 300$ pc and $T < 3 \times 10^4$ from all snapshots. The shaded region shows the 16th-84th percentiles. The two colors represent the two conditions for $\sigma_\perp$. The green and purple lines are reproduced from \cite{padovani_spectral_2021} and represent the values of s estimated from a combination of direct observations and modeling as detailed in \cite{orlando_imprints_2018} and \cite{padovani_cosmic-ray_2018} respectively. We have extrapolated the observed values at high energy using a constant slope for the sake of comparison (shown in the dashed lines). The black dashed line represents the input slope.}
    \label{fig:slope}
\end{figure}

In \autoref{fig:slope}, we show the slope of the CRE spectrum as a function of energy. Assuming that the spectrum between two consecutive bins can be approximated by a power law, we define the spectral slope as
\begin{equation}
    s = \frac{d{\rm log}\;j_e}{d{\rm log}\;E} \approx \frac{{\rm log}(j_{e,1}/j_{e,2})}{{\rm log}(E_1/E_2)}
    \label{eq:slope}
\end{equation}
where the subscripts represent consecutive CR bins. 

The solid, horizontal lines in \autoref{fig:slope} represent the median slope across all cells with $|z| < 300$ pc and $T < 3 \times 10^4$ from all snapshots. The shaded regions show the 16th-84th percentiles. We also include two lines representing models fit to the observed CRE spectrum from \cite{orlando_imprints_2018} and \cite{padovani_cosmic-ray_2018}. \cite{orlando_imprints_2018} use a CR propagation model to fit a combination of gamma-ray and radio emission along with direct CR observations. \cite{padovani_cosmic-ray_2018} use a multi-parameter fit to best match the direct observations from a combination of AMS and Voyager results. We see agreement between our simulations and these empirically-fit models in all energy bins. The CRE spectrum steepens with increasing energy due to changes in the dominant energy loss mechanisms and their varying dependence on CRE energy. As discussed in \autoref{sec:results_loss}, the contribution of synchrotron and IC increases with $E$, while Bremsstrahlung and ionization losses become less important. The former mechanisms have a stronger energy dependence ($\propto \gamma^2$) than the latter ($\propto \gamma \mathrm{ln}\gamma$ or $\mathrm{ln} \gamma$), which explains why the spectrum steepens at higher energies. The energy-dependent diffusion rate also contributes to steepening of the CRE spectrum. If advection and Alfvénic streaming were the dominant transport mechanism, we would not see any change in the spectral slope from the injected value where losses are negligible. This effect is more obvious in the proton spectrum where the energy loss mechanisms are not significant, and diffusion alone is responsible for energy-dependent steepening (Armillotta et al. 2025, accepted).

\begin{figure}
    \centering
	\includegraphics[scale = 0.57]{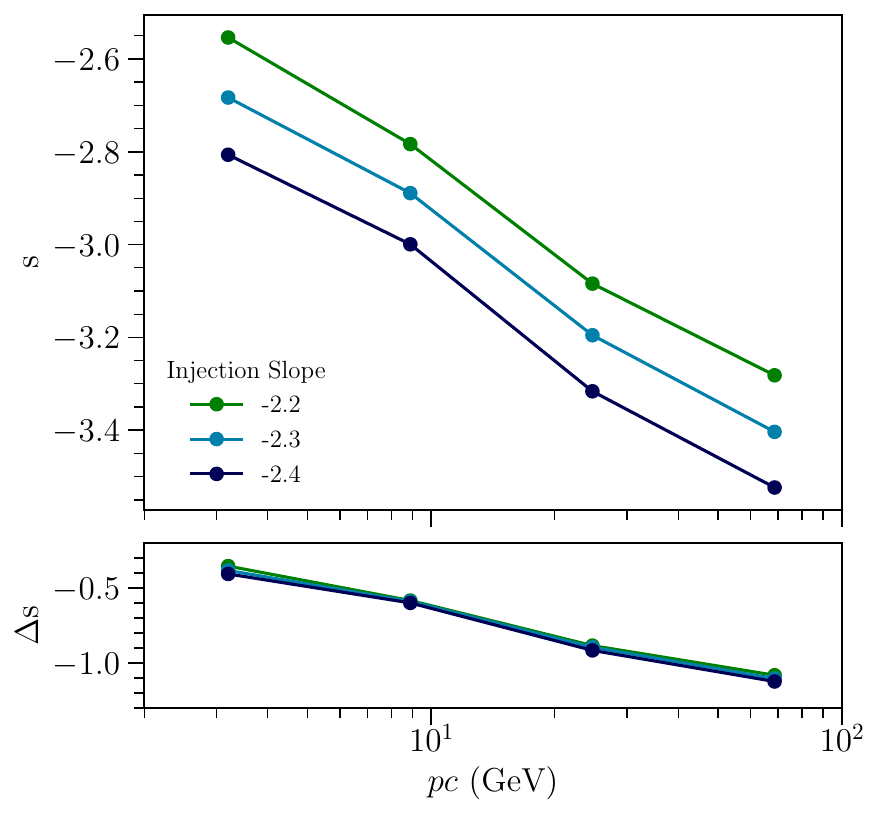}
    \caption{Power-law slope of the CRE spectrum (\autoref{eq:slope}) as a function of momentum for three choices of the injection spectrum. The points represent the median value across all cells with $|z| < 300$ pc and $T < 3 \times 10^4$ K. The lower panel shows the change in slope from the injected value.} 
    \label{fig:inject_slope}
\end{figure}

To determine the effect of the injection slope on the final CRE spectrum, we simulate the transport of CRs within the TIGRESS snapshot at $t = 214$~Myr using three different slopes of the injected distribution function: $\gamma_\mathrm{inj} = 4.2$, $4.3$ (our fiducial value), and $4.4$. These correspond to CRE spectral slopes: $s_\mathrm{inj} = -2.2$, $-2.3$, and $-2.4$. The evolved CR spectral slopes as a function of momentum from each model are shown in \autoref{fig:inject_slope}. This figure includes both the final value of $s$ (as in \autoref{fig:slope}), along with the change in slope, $\Delta s$, from the injected value. 

We find that the change in slope, $\Delta s$, does not vary significantly when the injection slope is modified. As noted in \autoref{sec:injection}, the CR energy density (\autoref{eq:energy}) is roughly linearly proportional to the injected spectrum except through effects from $\sigma_{\rm tot}$. Therefore, the final slope could be simply rescaled to a different choice of injection slope.  Based on  our physical result that $\Delta s$ varies with CRE momentum from $-0.4$ at 2 GeV to $-1.2$ at 100 GeV, we conclude that an injection slope of $s = -2.3$ is needed for a good match to the observed spectrum.

\subsection{Synchrotron Emission}
\label{sec:synch_results}

As a sample application for these simulations, we consider the primary observable for CREs, synchrotron radiation. Using the method described in \autoref{sec:synch}, we calculate the synchrotron emissivity in each snapshot as if an observer were looking vertically through the simulation box. We then integrate this emissivity to find the total synchrotron intensity. The total synchrotron intensity is dominated by emission near the midplane where CR feedback does not strongly affect ISM structure. Therefore, we can produce useful mock synchrotron observations even without self-consistent MHD and CR evolution. Each TIGRESS snapshot provides a mock observation of a 1 kpc square patch of a face-on galaxy. 

\begin{figure*}
    \centering
	\includegraphics[scale = 0.5]{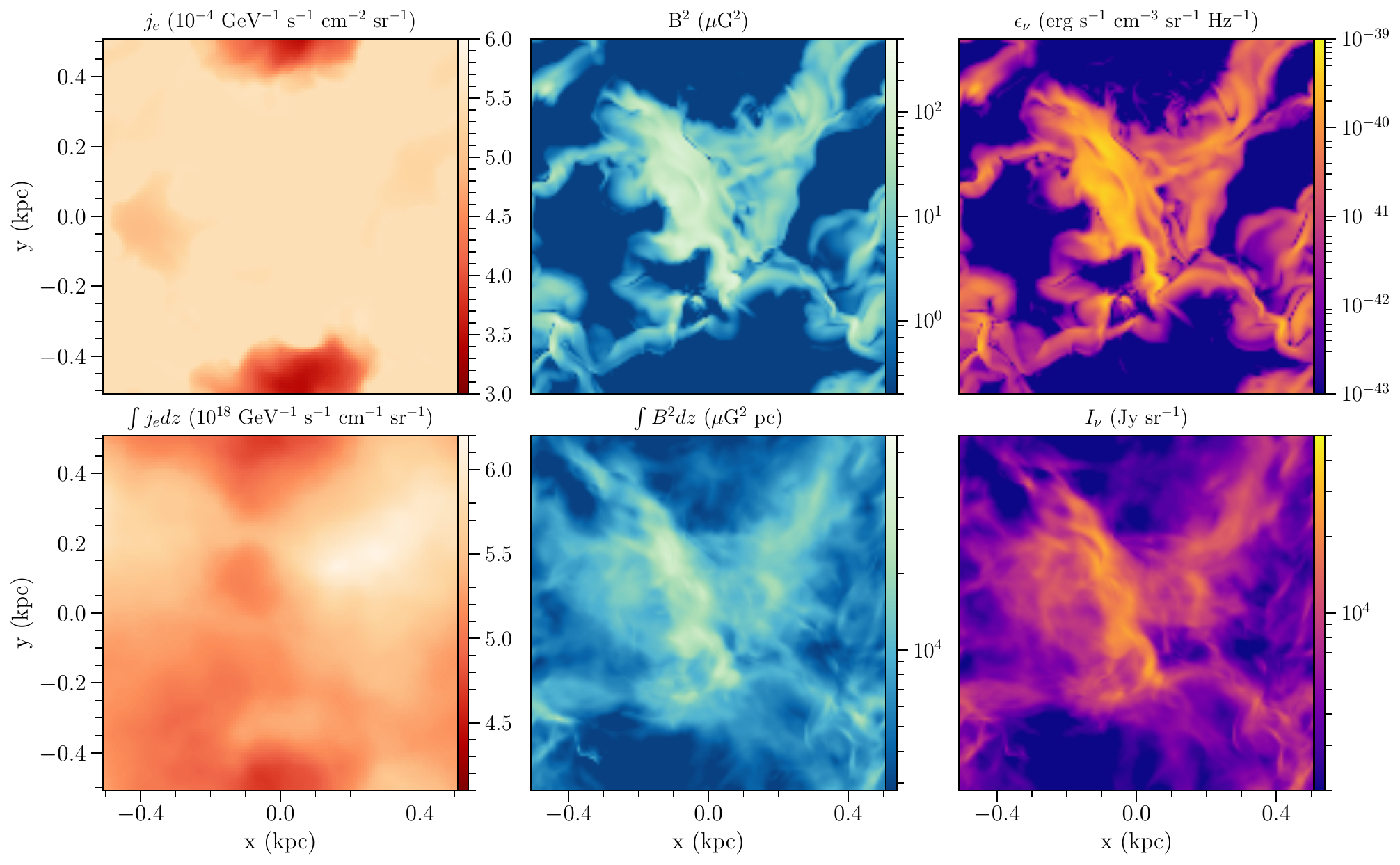}
    \caption{The upper panels represent slices at $z = 0$ of quantities relevant to synchrotron emission from the snapshot at $t=214$~Myr. From left to right, these are $j_e$, $B^2$, and synchrotron emissivity $\epsilon_\nu$. The lower panels represent these same quantities integrated vertically through the entire box. The lower right panel shows the synchrotron intensity $I_\nu$. The value of $j_e$ is for the 5 GeV electron energy bin. The synchrotron emissivity and intensity are evaluated at 1.5 GHz (L-band).}
    \label{fig:synch}
\end{figure*}

One example of the synchrotron emission at 1.5 GHz (L-band) is shown for the TIGRESS snapshot at $t = 214$~Myr in \autoref{fig:synch}. The upper row shows slices through $z=0$ of the CRE spectrum, $j_\mathrm{e}$, at $E = 5$~GeV, the square of the magnetic field strength $B^2$, and the synchrotron emissivity $\epsilon_\nu$. The vertically integrated values of each of these three quantities are shown in the second row. Both the single slice and vertical integral of $B^2$ show much more variation in the $x-y$ plane compared to $j_e$. Therefore, while the CRE spectrum is crucial for determining the overall magnitude of the synchrotron emission, the spatial distribution of synchrotron emission distribution is driven by that of the magnetic field.

Although the spatial distribution of synchrotron emission is almost independent of that of CREs, its distribution in frequency space is determined by the CRE spectral slope. 
Therefore, synchrotron observations at multiple frequencies can be used to estimate the underlying CRE spectral slope. Since the synchrotron spectrum nearly follows a power law, the synchrotron spectral slope can be computed from the ratio between the intensities at two different frequencies, $\nu_1$ and $\nu_2$, as
\begin{equation}
    \alpha = -\frac{{\rm log}(I_{\nu_1}/I_{\nu_2})}{{\rm log}(\nu_1/\nu_2)}\,.
\end{equation}
The corresponding CRE spectral slope can then be estimated as $s = -2\alpha - 1$ (see \autoref{sec:synch}).

Using our simulations, we can compare the true CRE spectral slope (as in \autoref{fig:slope}) to the slope that would be estimated from synchrotron observations. To do so, we generate maps of artificial synchrotron emission at eight frequencies corresponding to eight VLA radio bands (1.5, 3, 6, 10, 15, 22, 33, and 45 GHz)\footnote{https://science.nrao.edu/facilities/vla/docs/manuals/oss2013B
/performance/bands}. Between each pair of frequencies, we determine the radio spectral index and from this estimate the CRE spectral slope.

\begin{figure*}
    \centering
	\includegraphics[scale = 0.55]{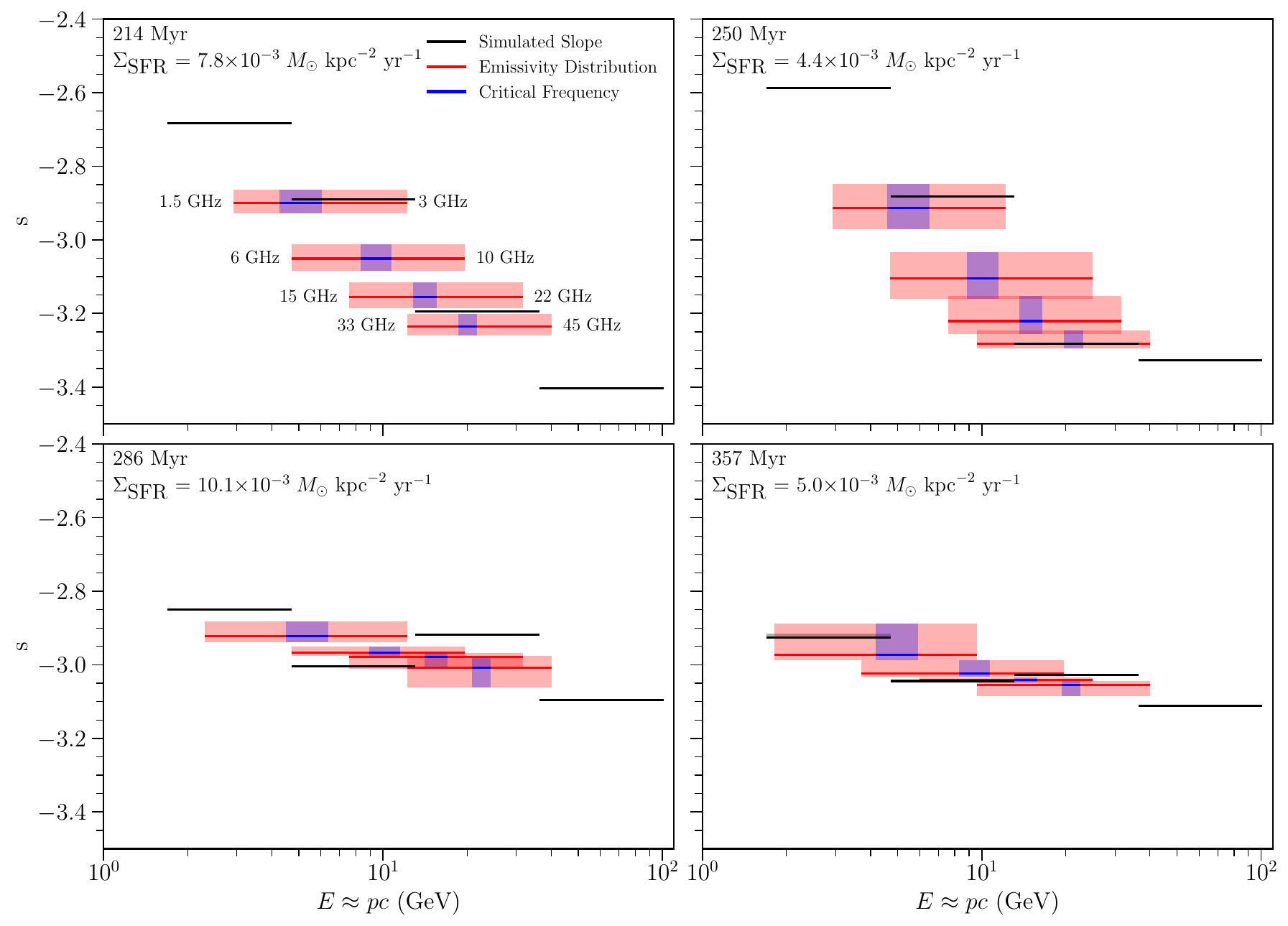}
    \caption{Comparison of the simulated CR power law slope to the value estimated from the mock synchrotron observations for four TIGRESS snapshots. The black bars represent the median power law slope of the simulated CR spectrum, evaluated over all cells within $z < |300|$ pc and with $T < 3\times10^4$ K (as in \autoref{fig:slope}). The red and blue bars represent the slope estimated from comparing the synchrotron intensities at the two frequencies listed to the side. The central line represents the median value, while the vertical extent shows the 16th-84th percentiles. The extension of the bars on the $x$-axis covers the range of energies of CREs contributing to the emission at those frequencies. The energy range shown in red is estimated from the integrand in \autoref{eq:eps_par} and \autoref{eq:eps_perp}, spanning the energies that represent the 16th-84th percentile of total emission. The blue bars span the critical energies corresponding to the two frequencies (\autoref{eq:ecrit}), assuming the magnetic field takes the equipartition value. Each panel is labeled with the snapshot time and the corresponding value of the SFR surface density.
    }
    \label{fig:synch_slope}
\end{figure*}

In \autoref{fig:synch_slope}, we compare the slopes of the simulated CR spectrum at different $E$ to the slopes estimated from the mock synchrotron observations. This is done for four of our snapshots. Each panel is labeled with the simulation time of the TIGRESS snapshot, along with the corresponding SFR surface density, $\Sigma_{\rm SFR}$. The true slopes of the underlying CRE spectrum (represented by black bars) are the median values calculated using \autoref{eq:slope}, including all cells within $z < |300|$ pc and with $T < 3\times10^4$ K, as in \autoref{fig:slope}. The majority of the variation in the slopes shown in \autoref{fig:slope} is due to differences between snapshots rather than between cells in one snapshot. Therefore, the error bars representing the 16th-84th percentile ranges of the slope are not visible in most of the panels of \autoref{fig:synch_slope}. These estimates for the CR slope represent the values that would be measured using direct detection, as we measure the CR spectrum in the solar neighborhood.

The red and blue bars in each panel represent the CRE slope estimated from the spectral index of the mock synchrotron emission. The dark central line within both the red and blue bars are the median value and the height of the bars shows the 16th-84th percentile range taking the distribution across all cells in the synchrotron map in the $x-y$ plane.
The locations of these bars along the $x$-axis represent the CRE energies which contribute to the synchrotron emission at each frequency. An individual CRE at a given energy will produce synchrotron emission at a range of frequencies, with a peak at a critical frequency. Therefore, synchrotron emission at a given frequency will be due to a population of CREs with a range of energies peaking at a critical CR energy. Any measurement of the synchrotron spectral index will probe the CR slope at this specific range of CR energy.

We estimate the CR energies which contribute to the synchrotron emission at each frequency in two different ways. The wider, red bars use the simulated values of synchrotron emissivity, $\epsilon_\nu$, to find which CRE energies are responsible for the majority of the synchrotron emission. To define this range, we find the energies which represent the 16-84th percentile around the peak of $\epsilon_\nu$ through \autoref{eq:eps_par} and \autoref{eq:eps_perp}. This estimate is not possible in reality, as only the integrated intensity rather than the emissivity is observable.

The narrower, blue bars represent an estimate based only on observable values. For a given frequency, $\nu$, the majority of emission will be due to CRs at energies of around
\begin{equation}
    E_{\rm crit} \approx \left(\frac{\nu}{16 \; \textrm{MHz}}\right)^{0.5}\left(\frac{B_\perp}{\mu \textrm{G}}\right)^{-0.5}
    \label{eq:ecrit}
\end{equation}
where $B_\perp$ is the component of the magnetic field perpendicular to the CR velocity \citep{beck_revised_2005}. The magnetic field is not known in observations, but a value can be obtained using the equipartition assumption. This is a commonly used way to treat the unknown magnetic field in extragalactic sources relying only on observed synchrotron emission. The primary assumption is that the CR energy density is in equipartition with the magnetic energy density. We refer to Appendix A of \citealt{beck_revised_2005} for the full derivation of $B_\perp$ under this assumption. The final expression is\begin{equation}
    \label{eq:equipartition}
    B_{\rm eq} = \left[ \frac{4 \pi (2\alpha + 1) (K_0 + 1)I_\nu E_{\rm p}^{1-2\alpha}(\nu/2c_1)^{\alpha}}{(2\alpha - 1)c_2(\alpha) l c_4(i)} \right]^\frac{1}{\alpha + 3}
\end{equation}
where $E_{\rm p}$ is the proton rest mass energy, $c_1$ is a numerical constant, and $c_2(\alpha)$ is a function dependent only on the spectral index. The correction for the inclination is given by $c_4(i)$. We use $c_4 = 1$ which is valid for a face-on view of a galaxy with a uniform magnetic field. $K_0$ represents the ratio in the number of CR protons to electrons at approximately 1 GeV. This is generally taken to be a factor of 100 and is approximately the same ratio observed in our evolved simulations. The length scale of synchrotron emission, $l$, must also be estimated and is typically on the order of a few kpc. We adopt $l= 1$ kpc, which is approximately twice the scale length of the emissivity. The choice of $K_0$ and $l$ do not have a large effect on the final value of $B_\perp$ due to the small power $1/(\alpha + 3)$ in \autoref{eq:equipartition}.

For each snapshot, we evaluate $B_\perp$ from the equipartition assumption using the mock synchrotron emission to compute $\alpha$. We then use this value to determine $E_{\rm crit}$. The blue bars in \autoref{fig:synch_slope} span the values of $E_{\rm crit}$ at each of the radio frequencies used for the estimate of the CR spectral index. 

We find that the value of the CR spectral index estimated from the synchrotron emission matches the directly measured values extremely well for each of the four snapshots shown in \autoref{fig:synch_slope}. The magnetic field is strongest, and the CR energy density is highest, at the midplane, so the synchrotron emission is dominated by the midplane gas. Therefore, the CR slope estimated from this emission would be expected to match the directly observed values at the midplane. The CR spectral index can be robustly recovered from synchrotron observations. We note that our mock radio observations do not include any thermal free-free emission, which would also be present especially at the higher frequencies we model. In true observations, this emission needs to be removed to learn about the underlying CR spectrum from the synchrotron emission. 

\section{Discussion}\label{sec:disc}

\subsection{Comparisons to other models}

Although many studies have simulated CR transport on galactic scales, only a limited number have modeled spectrally resolved CREs. As the total energy density of CREs is negligible compared to that of protons, they are not generally included in MHD simulations only concerned with the effects of CRs on ISM dynamics and galactic wind driving. CREs, however, represent a powerful observable, and simulating their transport on galactic scales is crucial to enable comparisons with observations (see \autoref{sec:obs}).

To date, other studies that have modeled the transport of spectrally resolved CREs in ISM/galaxy simulations include those by \cite{werhahn_cosmic_2021b, werhahn_cosmic_2021a} and \cite{hopkins_first_2022, hopkins_standard_2022}. \cite{werhahn_cosmic_2021b, werhahn_cosmic_2021a} model the transport of both protons and electrons in simulations of isolated Milky Way-like galaxies. These simulations have a mass resolution of $1.6 \times 10^{2-4}$ M$_\odot$ corresponding to a spatial resolution of approximately 20-90 pc at a number density of 1 cm$^{-3}$. They first run simulations where single-energy GeV CR protons evolve along with the background gas. These simulations treat CR transport in terms of advection and diffusion, while neglecting streaming, which we find to be an important transport process in the warm, ionized gas, where both advection and diffusion are limited \citep[][Armillotta et al.\ 2025, accepted]{armillotta_cosmic-ray_2024}. After self-consistently evolving GeV protons and MHD, they post-process their simulations to compute the distribution of spectrally resolved CR protons and electrons. To do so, in each computational cell, they solve the Fokker-Planck equation for the proton and electron distribution functions, assuming steady state, and neglecting streaming transport and adiabatic losses. Diffusion is parametrized by a spatially constant, energy-dependent diffusion coefficient based on observational constraints, $D = 10^{28} (E/3$ GeV)$^{0.5}$ cm$^{2}$ s$^{-1}$ \citep[see][]{evoli_ams-02_2020}. For the electrons, they account for energetic loss mechanisms such as synchrotron and IC. With this method, \cite{werhahn_cosmic_2021b, werhahn_cosmic_2021a} are able to reproduce the observed CR proton and electron spectra at a Galactocentric radius of $\simeq 8$~kpc.

The work by \cite{hopkins_first_2022} models energy-dependent transport of CREs, along with many other species, in cosmological zoom-in simulations. These models have a mass resolution of $\sim 10^4$ M$_\odot$ corresponding to a spatial resolution of approximately 75 pc at a number density of 1 cm$^{-3}$ \citep[][]{2019MNRAS.488.3716C, 2020MNRAS.492.3465H}. As in our model, they include CR transport via advection, diffusion, and streaming, as well as various energetic loss mechanisms, although they evolve the CR species self-consistently along with the background gas. Unlike our model, they also consider multiple methods of CR re-acceleration. Additionally, they employ a diffusion parameter that is dependent only on CR energy with $D \propto 10^{29} (R/1$ GV)$^{0.5}$ cm$^{2}$ s$^{-1}$, where $R = pc/q$ is the particle rigidity. The normalization of the diffusion coefficient is calibrated to reproduce observed spectra. We note that this diffusion coefficient is more than an order of magnitude higher than the value employed by \cite{werhahn_cosmic_2021b, werhahn_cosmic_2021a}. With this model, \cite{hopkins_first_2022} reproduce the observed CRE spectrum, along with the spectra of the CR protons and other heavier species.

In a subsequent work, \cite{hopkins_standard_2022} go beyond the assumption of spatially constant diffusion, and test a transport model with variable diffusion coefficient based on the predictions of the self-confinement scenario. With this model, however, they do not reproduce the observed CRE spectrum. A variety of reasons may contribute to the fact that the self-confinement model of \cite{hopkins_standard_2022} does not reproduce the observed CR spectrum, while our self-confinement model does.   As discussed in \cite{armillotta_cosmic-ray_2021} and \cite{armillotta_cosmic-ray_2024}, there are differences in the details of the CR implementation, as well as orders of magnitude difference in the mass resolution for hot gas.

It is encouraging, therefore, that our post-processing simulations using the self-confinement model are able to reproduce the observed CRE spectrum despite the simplifying assumptions made in the implementation of CR transport.
We find that  energy-dependent losses, in combination with energy-dependent CR diffusion using self-consistently determined scattering coefficients, and streaming and advection that are energy-independent (but strongly dependent on local multiphase gas properties), leads to energy-dependent steepening of the CRE spectrum. The CRE spectral slope obtained from our simulations is in good agreement  with direct observations in the solar neighborhood. The factor of two enhancement of the total CR energy density in our simulations compared to solar neighborhood observations could be attributed to a different energy injection rate per SN, or a difference in the SFR between the simulations and the Milky Way. 

\subsection{Applications to observations}
\label{sec:obs}

Direct measurement of the CRE spectrum is possible only in the solar neighborhood, and we have found our model to reproduce this result well. Our understanding of the CR distribution beyond the solar system relies on indirect observations, primarily through radio synchrotron emission. With both a spectrally resolved CRE population, and the magnetic field strength, we can produce synthetic synchrotron emission, as has been done using other CR transport simulations. \cite{ponnada_synchrotron_2023} produces synchrotron emission based on the simulations presented in \cite{hopkins_first_2022}, and \cite{werhahn_cosmic_2021b, werhahn_cosmic_2021a} include synthetic synchrotron radiation along with their other results. Additionally, \cite{chiu_simulating_2024} use the method described in \cite{werhahn_cosmic_2021a} to generate radio synchrotron spectra and polarization for a simulation of an edge-on galaxy. This mock synchrotron emission can be compared to radio observations to better understand the CR spatial distribution where direct detection of CRs is not possible.

Modeling of CR transport has been applied to many extragalactic radio observations \citep[e.g.][]{mulcahy_modelling_2016, schmidt_chang-es_2019, stein_chang-es_2023, heesen_diffusion_2023}. These models are generally greatly simplified compared to transport schemes in simulations due to a lack of knowledge about the local ISM properties necessary for more complex transport. The primary CR transport mechanism included in modeling observations is diffusion, generally with a diffusion coefficient that depends only on energy, although some models do also consider advection. These models may also include energetic losses through synchrotron and IC interactions.

We know from our CR simulations that the dominant transport mechanism varies between advection, streaming, and diffusion depending on the ISM phase \citep{armillotta_cosmic-ray_2024}. Also, the diffusion parameter can vary significantly based on local gas properties. Therefore, simplified models which ignore these mechanisms may not accurately reproduce the underlying CRE distribution. CREs with energies between approximately 100 MeV - 10 GeV are responsible for emission in the GHz range, and we know the CR spectral index varies significantly in this range, as observed directly and reproduced in models \citep{padovani_spectral_2021, bracco_new_2024}. Therefore, to model radio emission accurately, it is necessary to include an energy-dependent CRE spectral slope, which is not always done when interpreting observations \citep{padovani_spectral_2021}. 

In this work, we found that we are able to robustly recover the CR spectral index from synchrotron observations. Going forward we can test other assumptions made about the CR distribution from radio observations. For example, the equipartition assumption \citep[e.g.][]{beck_revised_2005} is often used to estimate the magnetic field strength in extragalactic sources \citep[e.g][]{beck_magnetic_2015, krause_chang-es_2018, mulcahy_investigation_2018}. This assumption may be approximately valid on kpc scales \citep[e.g.][]{stepanov_observational_2014}, but does not necessarily hold on smaller scales ($\sim$100 pc) where the CR distribution is nearly uniform spatially, but the magnetic field strength varies by orders of magnitude \citep[e.g.][]{seta_revisiting_2019}. Additionally, the equipartition may over- (or under-) estimate the underlying magnetic field strength on all scales \citep{dacunha_overestimation_2024}. With our simulations, we will be able to explore the accuracy of the equipartition assumption down to scales of order 10 pc, which can have applications to future, high-resolution radio surveys.

\section{Summary and Future Work}\label{sec:conc}

We have extended the model of CR transport presented in \cite{armillotta_cosmic-ray_2021} to include spectrally resolved CR protons and electrons between 1-100 GeV. This is done by post-processing TIGRESS simulations of the multiphase ISM which include accurate thermal structure, magnetic fields, and gas dynamics \citep{kim_three-phase_2017}, all of which are critical for CR transport. In particular, knowledge of local gas properties is necessary for determining the local rate of diffusion of the CRs, which is quite sensitive to the ionization state of gas since damping of small-scale magnetic fluctuations is very rapid in neutral gas.  Working within the self-confinement framework for CR transport, we self-consistently determine the scattering coefficient as a function of local gas and CR properties, based on the balance of streaming-driven Alfvén wave growth and regime-specific wave damping.

To model the evolution of the CRE spectrum accurately, we also include energy dependent losses. Through a combination of these losses and transport mediated by the self-consistently determined scattering coefficient, we see evolution of the CRE spectrum that is consistent with direct observations. Our results are robust to the choice of the perpendicular scattering rate and the initial CRE spectrum. We also find the simulated CR proton spectrum to be consistent with direct observations (Armillotta et al. 2025, accepted). For protons, losses are negligible so the final spectrum is determined primarily by the energy-dependent diffusion rate. To our knowledge, this represents the first model that both implements a scattering rate determined using the self-confinement model and reproduces the observed CR spectrum. A caveat, however, is that we have adopted a simplified post-processing model with independent CR momentum bins. It will therefore be valuable in the future to revisit this spectral analysis in a fully self-consistent model with live MHD evolution.

Because our simulations include a spectrally resolved CRE distribution along with realistic small-scale magnetic field structure, we can produce synthetic synchrotron emission. We find that the spatial structure of the synchrotron emission is dominated by the magnetic field distribution, but the CR spectrum determines the radio spectral index. We compare the directly observable CRE spectrum to that which can be extracted from radio observations and find the two to be consistent. This supports the observational method of indirectly determining the CR spectrum from radio observations. 

Another common usage of CREs is as a probe of the magnetic field strength through synchrotron observations, adopting an assumption of equipartition between the CR and magnetic energy density. While this approximation seems to hold on kpc scales, it does not necessarily hold true at smaller scales \citep[e.g.][]{seta_revisiting_2019}. We find the spatial distribution of CRs to be relatively flat compared to the magnetic field. In future work, we will assess the equipartition estimate of magnetic fields over a range of scales, down to our simulation resolution.

Our model of energy dependent CR transport presents many additional opportunities for future study. The discussion in this paper is limited to the CRE spectrum, but these simulations also include spectrally resolved CR protons as they are necessary for accurate determination of the scattering rate. We present models of the energy dependent CR proton distribution in Armillotta et al. (2025, accepted). Additionally, in this work we limit ourselves to solar neighborhood models of the ISM as a proof of concept, and also because for this environment we can compare to direct CRE observations. The transport model can, however, be applied to TIGRESS simulations representing many different galactic environments \citep[as in][]{armillotta_cosmic-ray_2022} to explore how the CRE spectrum may respond to environmental conditions in various regimes. The results presented in this paper focus on the midplane spectrum and face-on views of the simulated galaxy, and we do not consider the vertical variation in the CR spectrum. In future work, we may also consider how the CR spectrum changes with height, which will enable comparison to observations of edge-on galaxies \citep[e.g.][]{stein_chang-es_2023}. Finally, a limitation of the present CR models is that they are based on post-processing  multiphase ISM simulations followed by a brief period of active MHD. In the future, it will be of great interest to self-consistently include CR transport with live MHD and radiation.

\acknowledgments

We thank the referee for careful reading and constructive suggestions on the manuscript. Support for this work was provided by grant 510940 from the Simons Foundation to ECO and grant AST-2407119 from the NSF to LA and ECO. LA was supported in part by the INAF Astrophysical fellowship initiative.  EQ was supported in part by NSF AST grant 2107872 and by a Simons Investigator grant. 

\software{Athena \citep{stone_athena_2008, stone_gardiner2009}, Matplotlib \citep{Hunter:2007}, NumPy \citep{harris2020array}, SciPy \citep{2020SciPy-NMeth}, Astropy \citep{astropy:2013, astropy:2018, astropy:2022}, xarray \citep{hoyer2017xarray}}

\bibliography{references}

\appendix

\section{Effects of MHD relaxation}
\label{sec:app_mhd}

In \autoref{sec:transport}, we describe our two stage method for modeling CR transport. First, we evolve the CRs to a steady state while the MHD is frozen. Then, we allow for a brief period of MHD relaxation in which the gas can react to the presence of CRs. This short relaxation step reduces the unphysically large CR pressure gradients that form during the first post-processing stage.

\begin{figure*}[h]
    \centering
	\includegraphics[scale = 0.61]{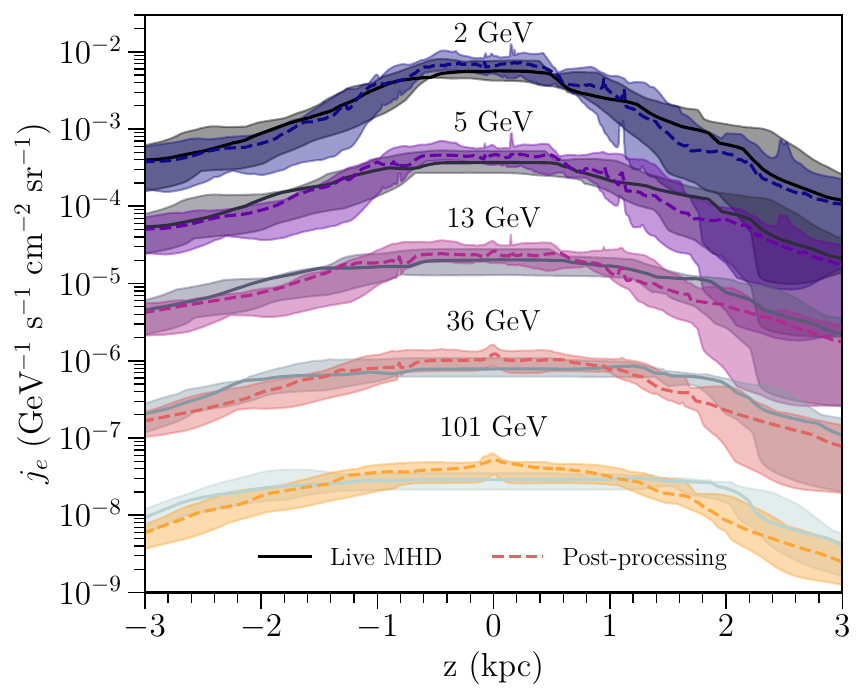}
    \includegraphics[scale = 0.61]{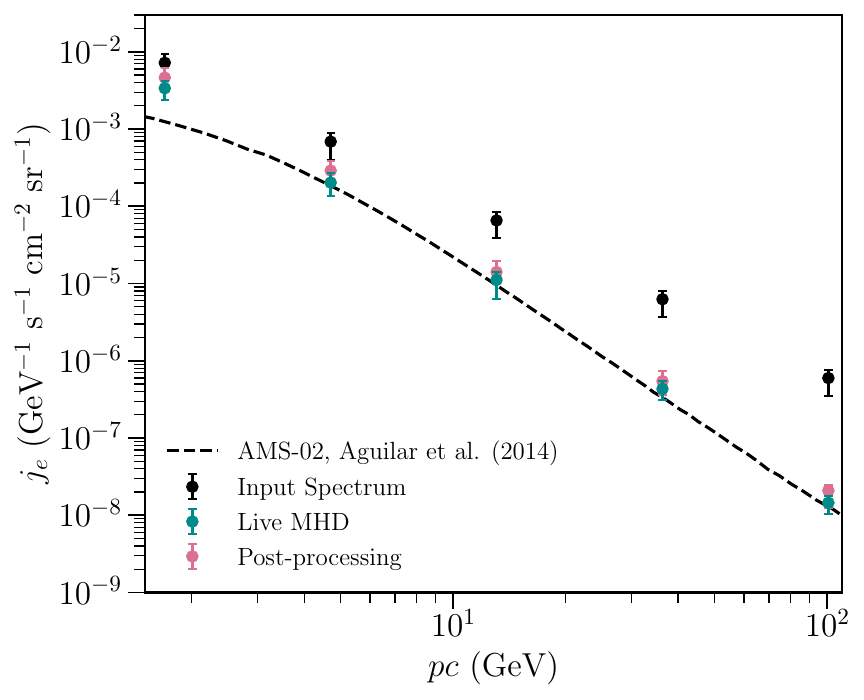}
    \caption{Comparisons of the CRE vertical profile and spectrum before and after live MHD relaxation. The left panel is analogous to \autoref{fig:zprof} and shows the horizontally and temporally averaged vertical profiles of the CRE spectrum, $j_e$, in each momentum bin. The shaded area covers the 16th and 84th percentiles from temporal variations while the central line represents the median value. The solid lines with a gray color scale represent the simulations after live MHD relaxation. The dashed lines with a purple color scale represent simulations with only CR post-processing, i.e. prior to the MHD relaxation step. The right panel is analogous to \autoref{fig:spectrum} and shows the spatially and temporally averaged value of $j_e$ in the warm gas ($T < 3 \times 10^4$~K) within the disk region ($\vert z \vert < 300$~pc). The points represent the median value across space and time, while the error bars show the 16th-84th percentiles. The blue and pink points represent the final CRE spectra with and without live MHD relaxation respectively. As in \autoref{fig:spectrum}, we include the input spectrum as black points, and reduce the normalization of all of the CRE spectra by a factor of two to more closely match the observed values from AMS \citep{aguilar_electron_2014}. All simulation snapshots have $\sigma_\perp = 10 \times \sigma_\parallel$.}
    \label{fig:MHD_comp}
\end{figure*}

The MHD relaxation step, because of its short duration, only has relatively local effects. We find that while the spatial distribution of the CRs becomes smoother, neither the vertical CRE profile nor the CRE spectrum change significantly after the relaxation step, as shown in \autoref{fig:MHD_comp}. We can see the overall normalization of the CRE spectrum decreases slightly after MHD relaxation. However, the shape of the vertical CRE profile and the CRE spectral slope are consistent before and after the live MHD stage.

\section{Proton losses}
\label{sec:app_p_losses}

\subsection{Pion production}

For CR protons with kinetic energies above $\sim 1$ GeV, the primary energy loss mechanism is pion production caused by elastic collisions with the surrounding atoms. Considering only interactions with hydrogen, the rate of energy loss due to pion production is given by \cite{krakau_pion_2015} to be
\begin{equation}
    \frac{dE}{dt} = 3.85 \times 10^{-16} n_{\textrm{H}}\bigg(\frac{E}{\textrm{GeV}}\bigg)^{1.28}\bigg(\frac{E}{\textrm{GeV}} + 200\bigg)^{-0.2}\textrm{ GeV s}^{-1}
\end{equation}
This energy loss rate is valid for kinetic energies greater than $\sim 10$ GeV where $E \approx E_k$. To extrapolate to lower kinetic energies, we define a continuous function with a constant slope below $E_k =$ 10 GeV  (Padovani priv. comm.)
\begin{equation}
\frac{dE}{dt} =
    \begin{cases}
        3.85 \times 10^{-16} n_{\textrm{H}}\big(\frac{E}{\textrm{GeV}}\big)^{1.28}\big(\frac{E}{\textrm{GeV}} + 200\big)^{-0.2}\textrm{ GeV s}^{-1} & \text{if } E_k > 10 \textrm{ GeV} \\
        \frac{dE}{dt}(E_k = 10 \textrm{ GeV}) \big(\frac{E_k}{10\textrm{ GeV}}\big)^{1.28}\textrm{ GeV s}^{-1} & \text{if } E_k < 10 \textrm{ GeV}
    \end{cases}
\end{equation}

The rate of energy loss is greater if we consider atoms other than hydrogen. If we assume the local ISM has a solar composition, we must multiply the energy loss by a factor of $\epsilon = 1.18$ \citep{padovani_impact_2020}. Writing the energy loss in terms of $\Lambda$ rather than $dE/dt$, we find
\begin{equation}
    \Lambda_{pion} =  1.18 \times \begin{cases}
        3.85 \times 10^{-16} \big(\frac{E}{\textrm{GeV}}\big)^{0.28}\big(\frac{E}{\textrm{GeV}} + 200\big)^{-0.2}\textrm{ cm}^{3}\textrm{ s}^{-1} & \text{if } E_k > 10 \textrm{ GeV} \\
        2.82 \times 10^{-15}\frac{1}{E/\textrm{GeV}}\big(\frac{E_k}{10\textrm{ GeV}}\big)^{1.28}\textrm{ cm}^{3}\textrm{ s}^{-1} & \text{if } E_k < 10 \textrm{ GeV}
    \end{cases}
\end{equation}

\subsection{Ionization}

At $E_k \sim 1$ GeV, the dominant loss mechanism for CRs in the neutral ISM is ionization of neutral hydrogen. The loss function is given by the Bethe-Bloch formula,
\begin{equation}
    L_{p,H}^{ion}(E) = \frac{4 \pi e^4}{m_e v^2}\bigg(\textrm{ln}\bigg(\frac{2 m_e v^2}{E_{ion}(1-\beta^2)}\bigg) - \beta^2\bigg) \textrm{ erg cm}^{2}
\end{equation}
\citep[e.g.][]{draine_physics_2011}. $E_{\rm ion}$ is the ionization energy of hydrogen, 13.6 eV, and $\beta = v/c$. The energy loss rate is then given by
\begin{equation}
    \frac{dE}{dt} = v x_{\mathrm{n}} n_H L_{p,H}^{ion}
\end{equation}
where we use the definition of the neutral number density given in \autoref{sec:ion_frac}. We again account for the solar composition by applying a multiplicative factor, $\epsilon = 1.1$ \citep{padovani_impact_2020}. Therefore, the total loss through ionization is given by
\begin{equation}
    \Lambda_{ion} =  1.1\frac{x_{\mathrm{n}}}{E} \frac{4 \pi e^4}{m_e v}\bigg(\textrm{ln}\bigg(\frac{2 m_e v^2}{E_{ion}(1-\beta^2)}\bigg) - \beta^2\bigg) \textrm{ cm}^{3}\textrm{ s}^{-1}
\end{equation} 

\subsection{Coulomb}

In ionized medium, the low energy CR protons primarily lose energy through Coulomb interactions. As in \cite{werhahn_cosmic_2021a}, this loss is given by \cite{gould1972energy} to be
\begin{equation}
    \frac{dE}{dt} = \frac{3 \sigma_T n_e m_e c^3}{2\beta} \bigg(\textrm{ln}\bigg(\frac{2 \gamma m_e c^2 \beta^2}{\hbar \omega_{pl}}\bigg) - \frac{\beta^2}{2}\bigg) \textrm{ erg s}^{-1} 
\end{equation}
where the plasma frequency, $\omega_{pl}$, is given by
\begin{equation}
    \omega_{pl} = \sqrt{\frac{4 \pi e^2 n_e}{m_e}} = \sqrt{\frac{4 \pi e^2 x_e n_H}{m_e}}
\end{equation}
and $\sigma_T = 6.65 \times 10^{-25}$ cm$^2$ is the Thomson cross-section. This gives a final value of,
\begin{equation}
    \Lambda_{coul} =\frac{x_e}{E}\frac{3 \sigma_T m_e c^3}{2\beta} \bigg(\textrm{ln}\bigg(\frac{2 \gamma m_e c^2 \beta^2}{\hbar \omega}\bigg) - \frac{\beta^2}{2}\bigg) \textrm{ cm}^{3}\textrm{ s}^{-1} = \frac{x_e}{E}\frac{4 \pi e^4}{m_e v} \bigg(\textrm{ln}\bigg(\frac{2 \gamma m_e v^2}{\hbar \omega}\bigg) - \frac{\beta^2}{2}\bigg) \textrm{ cm}^{3}\textrm{ s}^{-1}
\end{equation}

\section{Electron losses}
\label{sec:app_e_losses}

\subsection{Synchrotron}

CREs experience energy loss via interactions with the local magnetic field by emitting synchrotron radiation. The energy loss rate for one electron is
\begin{equation}
    \frac{dE}{dt} = \frac{c\sigma_T}{4\pi} (B \textrm{sin}(\theta))^2 \gamma^2 \textrm{ erg s}^{-1}
\end{equation}
\citep[e.g.][]{schlickeiser_cosmic_2002}. Here, $B$ is the magnitude of the magnetic field, $\theta$ is the pitch angle of the CR, and $\gamma$ is the Lorentz factor. Assuming an isotropic distribution of CRs, we can approximate
\begin{equation}
    \langle (B \textrm{sin}(\theta))^2 \rangle \approx \frac{2}{3} B^2
\end{equation}
(see e.g. \cite{torres_theoretical_2004}). Therefore, 
\begin{equation}
    \Lambda_{synch} = \frac{1}{6 \pi} \frac{\sigma_T}{m c} \gamma \frac{B^2}{n_H} \textrm{ cm}^3 \textrm{ s}^{-1}
\end{equation}

\subsection{Inverse Compton}

IC losses take an identical form to the synchrotron losses replacing the magnetic energy density with radiation energy density. The energy loss of one electron interacting with photons with radiation energy density, $w$, is \citep[e.g.][]{schlickeiser_cosmic_2002}
\begin{equation}
    \frac{dE}{dt} = \frac{4}{3} c\sigma_T w \gamma^2 \textrm{ erg s}^{-1}
\end{equation}
Therefore, 
\begin{equation}
    \Lambda_{IC} = \frac{4}{3} \frac{\sigma_T}{m c} \gamma \frac{w}{n_H} \textrm{ cm}^3 \textrm{ s}^{-1}
\end{equation}

We approximate the value of $w$ as a combination of the radiation from the CMB and starlight. The value of $w_{CMB}$ is a constant $\sim4.2\times10^{-13}$ erg/cm$^3$ \citep{draine_physics_2011}. The value of $w_{\rm star}$ is a combination of UV, optical, and IR emission. The TIGRESS-NCR simulations \citep{kim_introducing_2023} implement adaptive ray tracing to model the UV radiation field and have similar initial conditions to those of the TIGRESS models that we are post-processing. Therefore, we use the horizontally-averaged, vertical profile of the TIGRESS-NCR radiation field to estimate the value of $w_{UV}$. Additionally, we take a time average over a 200 Myr span of the TIGRESS-NCR simulation.

To account for optical emission, we use an estimate of the SED of a stellar population taken from \cite{kim_photochemistry_2023}. We multiply the value of $w_{UV}$ by the ratio of UV to optical emission in the SED to find $w_{opt}$.

The radiation energy density of IR emission must be found separately, as it will not be attenuated as the UV emission is. We approximate,
\begin{equation}
    w_{\rm IR}(x_0, y_0, z_0) = \frac{1}{4 \pi c} \iiint \frac{L_{IR}(x,y,z)dxdydz}{(x-x_0)^2 + (y-y_0)^2 + (z-z_0)^2}
\end{equation}
where $L$ is the IR luminosity at any point. For our solar neighborhood model, we take $x_0 = 8$ kpc and $y_0 = 0$ kpc. The radiation energy density is then only a function of $z$, as are $w_{UV}$ and $w_{opt}$. The IR luminosity of the Milky Way (representing emission from dust heated by starlight) is taken to follow a double exponential, 
\begin{equation}
    L(x,y,z) = A e^{-\sqrt{x^2 + y^2}/H_r}e^{-|z|/H_z}
\end{equation}
where we choose $H_r = 3$ kpc and $H_z = 200$ pc as the radial and vertical scale lengths. The normalization factor, A, is fixed such that the midplane value of $w_{IR}/w_{UV + opt}$ matches the observed value in the solar neighborhood \citep{draine_physics_2011}.

We rescale the total value of $w$ to the SFR of each post-processed TIGRESS snapshot by multiplying by the ratio of the SFR in the post-processed simulation to the average SFR in the TIGRESS-NCR model.

\subsection{Bremsstrahlung}

Low energy electrons (E $\lesssim$ 1 GeV) experience significant losses through bremsstrahlung interactions. The form of the energy loss can be divided into two limits, weak and strong shielding. In the weak shielding or completely ionized case, the expression for the energy loss is given by \cite{schlickeiser_cosmic_2002} to be
\begin{equation}
    \frac{d\gamma}{dt} = \frac{3\alpha c \sigma_T}{2\pi}\gamma \sum_Z n_Z Z(Z+1)\bigg(\textrm{ln}(\gamma) + \textrm{ln}(2) - \frac{1}{3}\bigg)
\end{equation}
where $\alpha$ is the fine structure constant. The sum included in this expression is over all ionized species with atomic number Z. Considering only hydrogen and helium, we have
\begin{equation}
    \sum_Z n_Z Z(Z+1) = 2 n_{\textrm{H}^+} + 6 (n_{\textrm{He}^{+}} + n_{\textrm{He}^{++}})
\end{equation}
Therefore, 
\begin{equation}
    \Lambda_{\rm BS, i} = \frac{3\alpha c \sigma_T}{\pi}\bigg(x_{\rm{H}^+}  + 3 x_{\rm{He}^+}\bigg)\bigg(\textrm{ln}(\gamma) + \textrm{ln}(2) - \frac{1}{3}\bigg)
\end{equation}
where $x_{\rm{H}^+} = \frac{n_{\rm{H}^+}}{n_H}$ and $x_{\textrm{He}^{+}} = \frac{n_{\textrm{He}^{+}} + n_{\textrm{He}^{++}}}{n_H}$. These values are determined as in \autoref{sec:ion_frac}.

In the neutral or strong shielding limit, the expression for bremsstrahlung losses is instead given by \cite{schlickeiser_cosmic_2002} to be
\begin{equation}
    \frac{d\gamma}{dt} = \frac{3.9 \alpha c \phi_{\rm{HI}}^{s-s} \sigma_T}{8\pi}\gamma (n_{\rm{HI}} + 2 n_{\rm{H2}})
\end{equation}
where $\phi_{\rm{HI}}^{s-s} \approx 45$ is the scattering function. This gives,
\begin{equation}
    \Lambda_{\rm BS, n} = \frac{3.9 \alpha c \phi_{\rm{HI}}^{s-s} \sigma_T}{8\pi} x_{\mathrm{n}}
\end{equation}
The total bremsstrahlung loss expression is then
\begin{equation}
\Lambda_{\rm BS} = \frac{\alpha c \sigma_T}{\pi} \bigg(3\big(x_{\rm{H}^+}  + 3 x_{\rm{He}^+}\big)\Big(\textrm{ln}(\gamma) + \textrm{ln}(2) - \frac{1}{3}\Big) + \frac{3.9\phi_{\rm{HI}}^{s-s}}{8} x_{\mathrm{n}} \bigg)
\end{equation}

\subsection{Ionization and Coulomb}

At the lowest energies, ionization and Coulomb interactions are the dominant sources of energy losses in CREs. From \cite{schlickeiser_cosmic_2002}, the rate of energy loss due to ionization of neutral hydrogen and helium is
\begin{equation}
    \frac{d\gamma}{dt} = 2.7 c \sigma_T (6.85 + \textrm{ln}(\gamma)) (n_{\rm{HI}} + 2 n_{\rm{H2}})
\end{equation}
Therefore,
\begin{equation}
    \Lambda_{\rm ion} =  2.7 c \sigma_T \frac{6.85 + \textrm{ln}(\gamma)}{\gamma} x_{\mathrm{n}}
\end{equation}
In ionized medium, Coulomb interactions are more important than ionization losses. Interactions with free electrons results in energy losses with the form \citep{schlickeiser_cosmic_2002}
\begin{equation}
    \frac{d\gamma}{dt} = \frac{3}{4} c \sigma_T n_e \big(74.3 + \textrm{ln}(\gamma) - \textrm{ln}(n_e)\big)
\end{equation}
Therefore, 
\begin{equation}
    \Lambda_{\rm Coulomb} = \frac{3}{4} c \sigma_T \frac{74.3 + \textrm{ln}(\gamma) - \textrm{ln}(n_e)}{\gamma}  x_e
\end{equation}
The total energy loss from these two terms is then
\begin{equation}
    \Lambda_{\rm ion + Coulomb} = \frac{3}{4} \frac{c \sigma_T}{\gamma} \bigg(3.6 \big(6.85 + \textrm{ln}(\gamma)\big) x_{\mathrm{n}} + \big(74.3 + \textrm{ln}(\gamma) - \textrm{ln}(n_e)\big)  x_e\bigg)
\end{equation}

\end{document}